\documentclass[aps,prb,twocolumn,superscriptaddress,longbibliography]{revtex4-2}
\usepackage[pdftex]{graphicx}
\usepackage{mmap}
\usepackage{amssymb}
\usepackage{amsmath}
\usepackage{color}
\usepackage[usenames,dvipsnames]{xcolor}
\usepackage{hyperref}
\hypersetup{colorlinks=true,linkcolor=blue,urlcolor=blue,citecolor=blue}

\begin{document}

\title{Hybridization of edge modes with substrate phonons}

\author{Azat F. Aminov}
\affiliation{National Research University Higher School of Economics, 100100 Moscow, Russia}
\affiliation{Institute of Microelectronics Technology and High Purity Materials, Russian Academy of Sciences, Chernogolovka 142432, Russia}
\author{Alexey A. Sokolik}
\email{asokolik@hse.ru}
\affiliation{Institute for Spectroscopy, Russian Academy of Sciences, 108840 Troitsk, Moscow, Russia}
\affiliation{National Research University Higher School of Economics, 100100 Moscow, Russia}

\begin{abstract}
Surface plasmons or phonons propagating on a two-dimensional (2D) material can exhibit coupling with resonant excitations in its substrate, and the resulting coupled modes were extensively studied. Similar coupling of edge modes propagating along a boundary of 2D material with the substrate excitations remains unexplored. This paper aims to bridge this gap by investigating the edge-substrate hybrid modes using the exact Wiener-Hopf analytical approach. We analyze dispersions, decay rate, and confinement of such hybrid modes both without magnetic field (plasmon-phonon and phonon-phonon hybrid modes) and in quantizing magnetic field where edge magnetoplasmons hybridize with substrate phonons. The hybrid modes are predicted to occur in THz and far-IR ranges for several combinations of quasi-2D materials (single- and bilayer graphene, quantum wells, thin-film semiconductors) and substrates with polar-phonon resonances.
\end{abstract}

\maketitle

\textbf{Introduction.} Light localization at the nanoscale, the core advantage of plasmonics, opens the way to various real-life applications, including sub-Doppler microscopy, molecular sensing, information transmitting, etc \cite{Huang_2016,SINGH2023100033}. In particular, with the aim of utilizing plasmons for molecular and substance detection, a considerable body of research has been undertaken to investigate coupling between plasmons and phonons \cite{Huang_2016}. To this day, several experiments have touched upon this topic: the interplay of surface plasmons on graphene with phonons in SiO$_{2}$ \cite{Luxmoore_2014, Yan_2013}, SiC \cite{Koch_2010}, and hBN \cite{Huang_2023, Wehmeier_2024}; the existence of plasmon-phonons in bilayer graphene \cite{Yan_2014} and on the surface of several semiconductors (InSb \cite{Bryxin_1972}, SiC \cite{Greffet_2002}, Al$_x$Ga$_{1-x}$As \cite{Po_ela_2014}, GaN \cite{Melentev_2016}). %Adato_2009, Rodrigo_2015 убрал

Edge plasmons emerge at the boundary between a thin bounded material and a vacuum \cite{Mast_1985, Volkov1988EdgeM}. They exhibit stronger field confinement in comparison with the surface plasmons, amplifying light-matter interaction and promising improved spatial resolution for sensing applications. Edge magnetoplasmons (EMP) arising in a magnetic field \cite{Mast_1985,Volkov1988EdgeM,MikhailovChapter1E, Sokolik_2024}, in addition to the standard advantages of plasmons, allow for even more tunability of their properties \cite{SINGH2023100033}, and demonstrate weaker damping \cite{Yan_2012}, stronger field confinement \cite{MikhailovChapter1E}, and topological protection \cite{Jin_2016}. Additionally, they can unidirectionally propagate along complex edge geometries, enabling designs of plasmonic circuitry \cite{Mahoney_2017}. Importantly, in some topologically non-trivial materials, like metamaterials \cite{Xiao_2024} or quantum anomalous Hall insulators \cite{Mahoney_2017} and metals \cite{Song_2016}, EMPs can exist without magnetic field.

While resonant coupling of surface plasmons with phonons was previously extensively investigated \cite{Huang_2016, Singh_2023}, no such consideration has been bestowed upon the edge plasmons. In this study, we present the first, to our knowledge, investigation of edge mode hybridization with substrate excitations, which can be phonons, plasmons, excitons, or other optically active modes. Using the exact Wiener-Hopf method, we find the dispersion law for edge modes with nonzero damping and show how it is affected by substrate resonances. Besides, we construct analytical approximations and show that the anti-crossed dispersion picture of the predicted hybrid mode branches can be understood in terms of a coupled-oscillators model.

For brevity, the novel hybrid edge-substrate modes will be referred to in the following as simply edge modes, or as EMPs in the presence of external magnetic field.

\begin{figure}[t]
   \centering   \includegraphics[width=1\linewidth]{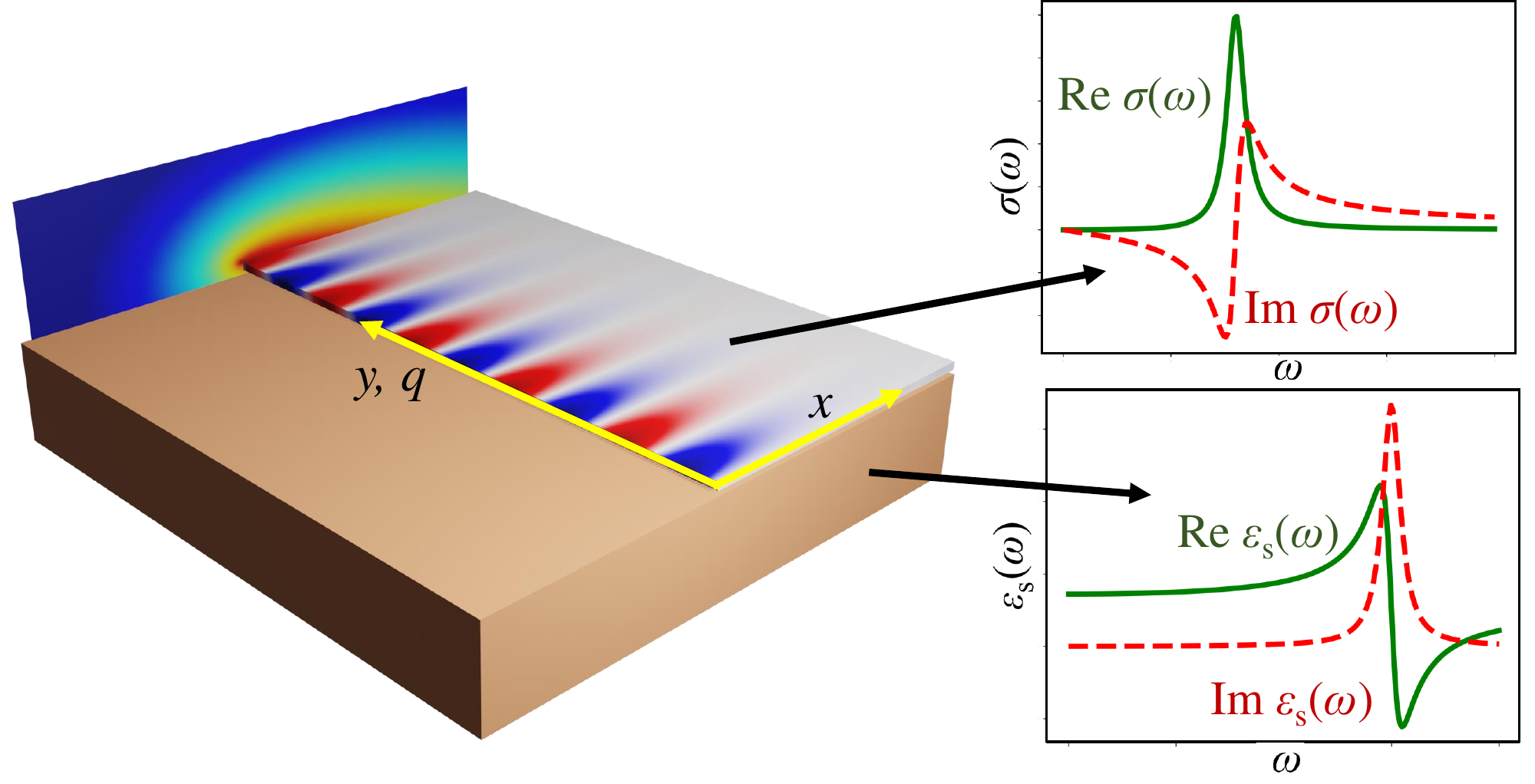}
   \caption{Depiction of the system: thin-film or 2D material with a sharp edge (half-plane $x>0$, $z=0$), having the surface conductivity $\sigma(\omega)$, is put on a substrate with the dielectric function $\varepsilon_\mathrm{s}(\omega)$. Both $\sigma(\omega)$ and $\varepsilon_\mathrm{s}(\omega)$ have resonances at close frequencies. Edge modes with the wavevector $q$ are moving along the $y$ axis, and  distribution of their oscillating scalar potential is depicted in the background.}
   \label{fig:Scheme}
\end{figure}

\textbf{System overview and substrate resonances.} We consider systems, schematically depicted in Fig.~\ref{fig:Scheme}. Thin-film or 2D material bounded by a straight sharp edge is placed on a bulk substrate with the dielectric function $\varepsilon_\mathrm{s}(\omega)$, and may be optionally coated by another thick material with dielectric function $\varepsilon_{\mathrm{c}}(\omega)$. Retardation effects are negligible far away from the light cone $2\pi/q\ll\lambda_{0}$, when the wavelength of the mode along the edge is larger than the excitation wavelength in vacuum $\lambda_{0}=2\pi c/\omega$. For frequencies near 10 THz, this condition translates to $q\gg 0.2$~$\mu$m$^{-1}$. In the non-retarded limit $c\to\infty$ the modes emerging in 2D material between two dielectrics are affected only by the mean dielectric function $\varepsilon=\tfrac12\left( \varepsilon_\mathrm{s}+\varepsilon_\mathrm{c}\right)$ (see details in the Supplement 1). This function, whose resonances are given by optically active excitations such as phonons or excitons, can be approximated by the Lorentz formula
\begin{equation}\label{varepsilon}
    \varepsilon(\omega) =  \varepsilon_{\infty}-\frac{1}{2}\sum_j\frac{f_{j}\omega_{\mathrm{P},j}^{2}}{\omega^{2} - \omega_{\mathrm{P},j}^{2} + i\omega\Gamma_{\mathrm{P},j} } ,
\end{equation}
where $\varepsilon_{\infty}=\tfrac12(\varepsilon_{\mathrm{s},\infty}+\varepsilon_{\mathrm{c},\infty})$ is the mean high-frequency dielectric constant of the surrounding, $\omega_{\mathrm{P},j}, \Gamma_{\mathrm{P},j}, f_{j}$ are resonant frequency, damping rate, and oscillator strength of the $j$th mode in substrate or coating materials. In this paper, we focus on phonon resonances, so $\omega_{\mathrm{P},j}$ is the frequency of transverse optical phonon. To ease visualization and excitation of plasmons, the coating is often not applied ($\varepsilon_\mathrm{c}=1$), and we also omit it. There exist several ways to determine oscillator strengths from the known functions $\varepsilon_{\mathrm{c,s}}(\omega)$ employing the Lyddane-Sachs-Teller or Cochran-Cowley-Kurosawa relations \cite{Fischetti_2001, Chaves_1973, Takahashi_1975}. The methods we use to extract $f_{j}$ for our calculations are presented in Supplement 1.

At the bare boundary between the substrate and vacuum, the surface phonons (or plasmons for metallic substrate) emerge, their frequencies being given by $\varepsilon(\omega)=0$ in the non-retarded limit. We denote by $\omega_{\mathrm{SP},j}$ the frequencies of such surface phonons (counterpart of plasma frequency). If the substrate resonances are well separated and the damping is weak, they are given by the formula
\begin{equation}\label{w_SP_and_Omega_P}
      \omega_{\mathrm{SP},j} = \omega_{\mathrm{P},j} \sqrt{1 + f_{j}/2\varepsilon_{\infty}},
\end{equation}
from which the important relation $\omega_{\mathrm{SP},j}>\omega_{\mathrm{P},j}$ follows.

\textbf{Edge modes.} Throughout this section we will consider thin-film or 2D materials with a sharp edge and pronounced isotropic resonance in the  spectrum of their surface conductivity $\sigma_{xx,yy}(\omega)$. We limit ourselves to the situation when only one resonance is prominent in the spectral range we are interested in (THz to far-IR), so we may use the general formula
\begin{equation}\label{sigma}
    \sigma_{xx,yy}(\omega)=\frac{iD}{\pi}\frac{\omega} {\omega^{2}-\Omega^{2}+i\omega\Gamma} ,\quad \sigma_{xy,yx}(\omega)=0.
\end{equation}
Here $D=2\int_{0}^{\infty} \mathrm{Re}\,\sigma_{xx} (\omega)d\omega$ is the weight (with dimension $\mathrm{cm/s}^2$) of the resonance at frequency $\Omega$ with the broadening $\Gamma$; $\Omega$ equals the frequency of a transverse optical phonon $\omega_{\mathrm{P},j}$ for polar-phonon thin-film material, or $\Omega=0$ for the Drude model of a 2D conductor. 

\begin{figure}[t]
  \includegraphics[width=\linewidth]{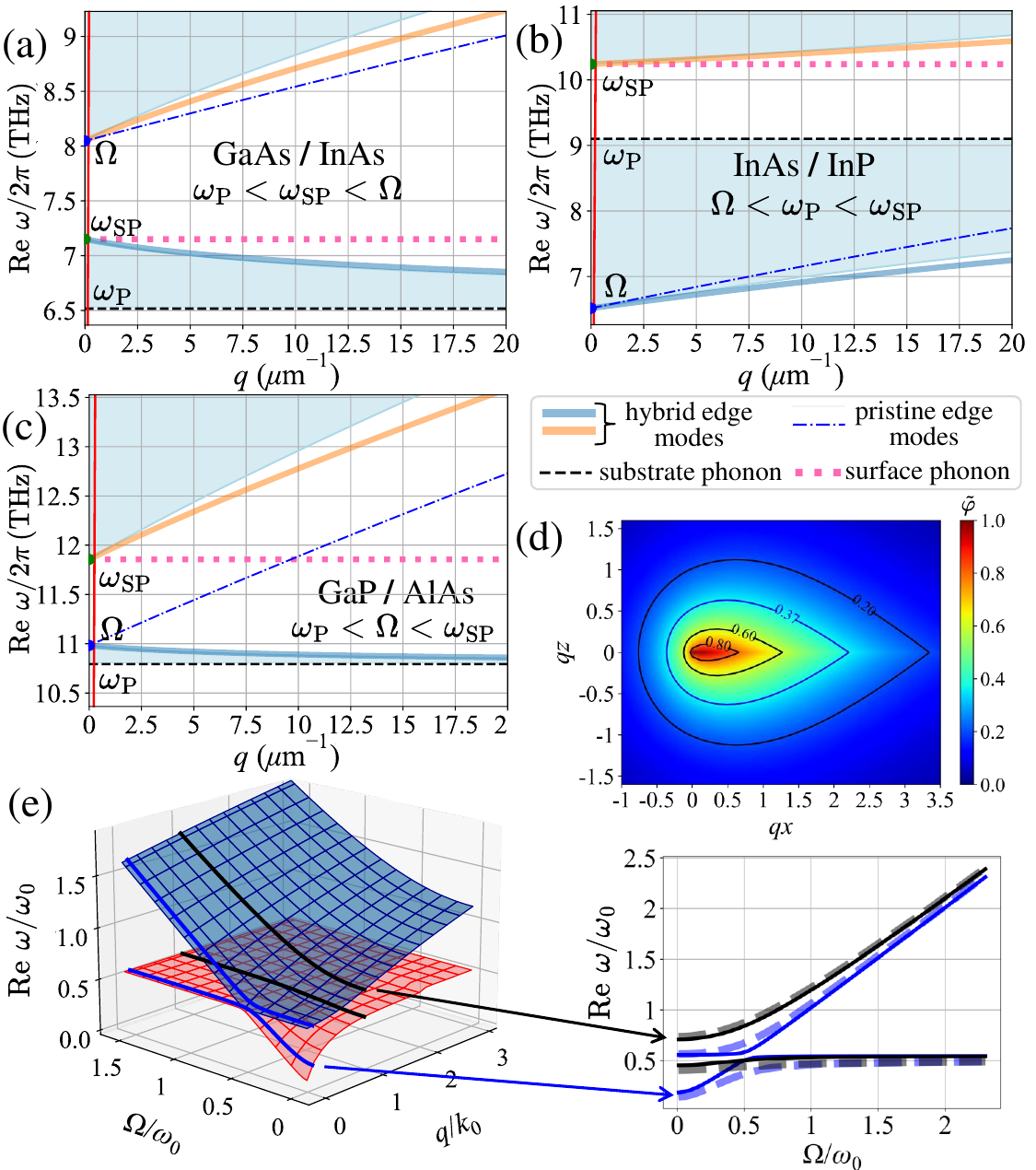}
  \caption{(a,b,c) Dispersions of hybrid edge modes emerging in thin polar films on polar substrates, compared with the surface ($\omega_{\mathrm{SP}}$), bulk ($\omega_\mathrm{P}$), and 2D material ($\Omega$) phonon frequencies. Blue shaded areas are continua of surface modes, and red lines denote the light cone. The chosen pairs of thin-film and substrate materials demonstrate three possible mutual arrangements of the frequencies $\omega_{\mathrm{P}}$, $\omega_{\mathrm{SP}}$, $\Omega$. (d) Rescaled spatial profile of scalar potential for the edge modes.
  (e) Dispersions of edge modes exhibiting anti-crossing at varying $\Omega$, which is well approximated by the coupled-oscillator model, as shown by dashed lines at two selected wavevectors $q$ at the right (calculation parameters are $ \sqrt{f/4\varepsilon_{\infty}}=0.3$, $\omega_{\mathrm{P}}/\omega_{0}=0.5$, where $\omega_{0}=4D/c$, $k_{0}=c/\omega_{0}$).}
    \label{fig:Field_B=0}
\end{figure}

Maxwell equations describing electromagnetic field and charge oscillations in the system depicted in Fig.~\ref{fig:Scheme} can be solved analytically using the Wiener-Hopf method \cite{Volkov1988EdgeM,Sokolik_2021}. It provides the dispersion equation for edge modes with frequency $\omega$ and wavevector $q$ along the edge,
\begin{equation}\label{dispersion_B=0}
    \frac{4\pi q\sigma_{xx}(\omega)}{i\varepsilon(\omega)\omega}=\eta_{0},
\end{equation} 
where $\eta_{0}\approx2.43$ \cite{Volkov1988EdgeM}. Thanks to essentially resonant frequency dependence of $\varepsilon(\omega)$, solution of Eq.~(\ref{dispersion_B=0}) provides dispersions of the edge-substrate hybrid modes instead of pristine edge modes arising in the case $\varepsilon=\mathrm{const}$. The examples, calculated for 2D materials and substrates with the phonon resonances, are plotted in Fig.~\ref{fig:Field_B=0}(a,b,c). Hybrid plasmon-phonon modes, arising in the case of 2D metal, originate from the same near-field electromagnetic coupling and are described by the same equations (\ref{varepsilon}) (\ref{dispersion_B=0}), but with the Drude conductivity (\ref{sigma}) at $\Omega=0$. Example of such modes is considered in Supplement 1.

Besides the edge modes near the boundary of a half-plane, there are surface modes propagating along the bulk of 2D material. The frequency of these modes, whose wavevector shares the same $y$-projection $q$ as the edge modes but additionally has the $x$-component $k$, can be determined from the standard equation (see Eq.~(14) in Ref.~\cite{Sokolik_2021} and Supplement 1 for details):
\begin{equation}\label{dispersion_2D}
    \frac{4\pi \sqrt{k^{2}+q^{2}} \sigma_{xx}(\omega)}{i\varepsilon(\omega)\omega}=2.
\end{equation}
Varying $k$ from 0 to $\infty$, we obtain the whole continuum of surface modes at given $q$, which does not intersect the edge mode dispersion.

\begin{figure*}[t]
\begin{center}
\includegraphics[width=0.85\textwidth]{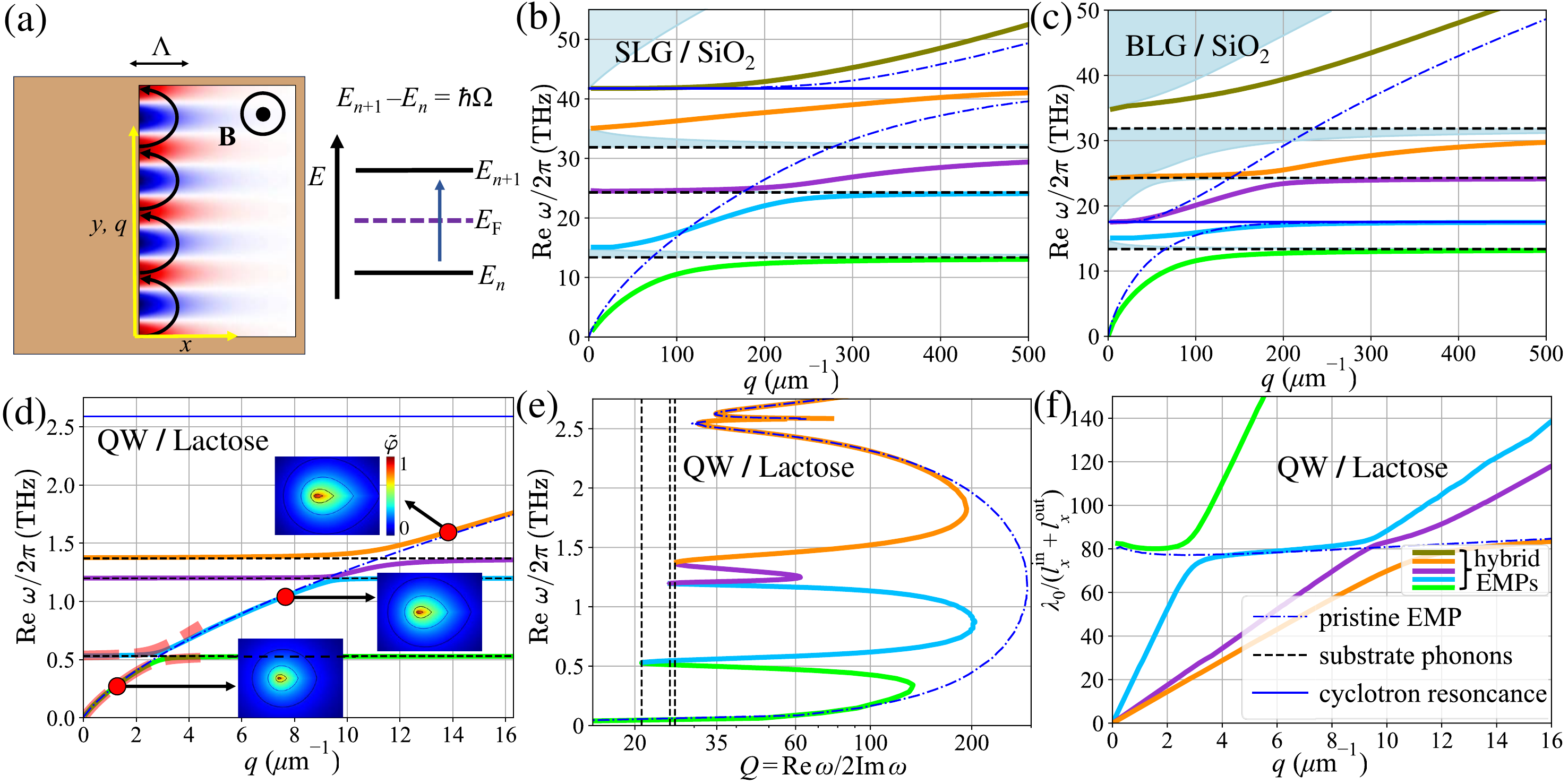}
  \end{center}
  \caption{(a) Edge magnetoplasmons arise in magnetic field applied perpendicularly to laterally bounded 2D material, whose optical inter-Landau-level transition across $E_\mathrm{F}$ gives rise to the cyclotron resonance at $\Omega$. (b,c,d) EMP dispersions for bounded QW, SLG and BLG on crystalline lactose or SiO$_{2}$ substrates with polar-phonon resonances. Shaded regions are continua of surface magnetoplasmon modes, and notation of other dispersion curves are given on panel (f). Thick dashed lines in (d) corresponds to the coupled-oscillator model (\ref{matrix}), and insets show the examples of potential profiles in space. (e) Quality factor of EMPs, phonons, and pristine EMP. (f) Field confinement factors for each EMP mode in (d) shown by the same color coding.}
    \label{fig:Magnetoplasmons}
\end{figure*}

Let us briefly analyze dispersions of edge modes given by the real part $\mathrm{Re}\,\omega(q)$ of the frequencies obtained from Eq.~(\ref{dispersion_B=0}) for a single substrate resonance (so the index $j$ is omitted). This equation has two positive solutions $\omega_{1,2}(q)$, and two reciprocal solutions with the opposite sign $\omega_{3,4}(q)=-\omega_{1,2}(q)$. Notice that at $q=0$ one dispersion branch $\omega_1(q)$ originates from $\omega_{\mathrm{SP}}$ (where $\varepsilon(\omega) = 0$), and the second branch  $\omega_{2}(q)$ originates from $\Omega$ (where $\sigma_{xx}(\omega)\to \infty$). Since $\omega_\mathrm{SP}>\omega_\mathrm{P}$, three arrangements of the frequencies $\omega_\mathrm{P}$, $\omega_\mathrm{SP}$, $\Omega$ are possible. Two of them arise when the resonance frequency of the 2D material is the highest, $\omega_\mathrm{P}<\omega_\mathrm{SP}<\Omega$ (Fig.~\ref{fig:Field_B=0}(a)), and when it is the lowest, $\Omega<\omega_\mathrm{P}<\omega_\mathrm{SP}$ (Fig.~\ref{fig:Field_B=0}(b)). The remaining arrangement $\omega_\mathrm{P}<\Omega<\omega_\mathrm{SP}$ (Fig.~\ref{fig:Field_B=0}(c)) requires specific tuning of material parameters, because the resonance $\Omega$ of the 2D material must fit in the interval between the closely spaced resonant frequencies $\omega_\mathrm{P}$ and $\omega_\mathrm{SP}$ (\ref{w_SP_and_Omega_P}) of the substrate. We have found materials for all these three situations and calculated hybrid edge mode dispersions (see the calculation parameters in Supplement 1).

Distribution of scalar potential amplitude $\phi(x,z)$ of the edge mode in space can be calculated using Eqs.~(C2)--(C3) in Ref.~\cite{Sokolik_2021}. Remarkably, under the assumptions we are working with (non-retarded limit, materials much thinner than excitation wavelength) these expressions depend only on $qx,qz$ and do not include explicit $\omega$ or mode branch dependence, thus the rescaled spatial distribution of the electromagnetic field, represented by $\tilde{\varphi}(qx,qz)=\varphi(x,z)/\varphi(0,0)$, is invariant and shown by Fig.~\ref{fig:Field_B=0}(d). The field confinement near the edge can be described by the surface where $\varphi(x,z)=\varphi(0,0)/\mathfrak{e}$ ($\mathfrak{e}\approx2.718$), which is shown by blue line. The corresponding confinement lengths along the $x$ and $z$ axes, at which the field decreases $\mathfrak{e}$ times, are $l_i=C_i/q$, where $C_i=2.23$, $0.418$, and $0.603$ for $l^{\mathrm{in}}_{x}$, $l^{\mathrm{out}}_{x}$, and $l_{z}$, respectively, and in (out) corresponds to positive (negative) $x$ direction. The confinement factors, showing how strongly the field is localized near the edge with respect to $\lambda_0$, are given by $\lambda_{0}/l_i= 2\pi cq/\omega  C_i$. Since $\lim_{q\to0}\omega\ne0$, these factors tend to zero at small wavevectors $q$. We have also calculated quality factors of edge modes and coefficients of light reflection from the structure, demonstrating that edge mode frequencies mainly reside inside the substrate Reststrahlenbands (see Supplement 1).

The hybrid mode dispersion may be qualitatively described in terms of the coupled-oscillator model \cite{Wan_2016} as near-resonant coupling of pristine edge mode with the substrate resonance. This model results from the approximation $\omega\varepsilon(\omega)\approx\omega_{\mathrm{pr}}(q)\varepsilon_{\infty}$ in Eq.~(\ref{dispersion_B=0}), where $\omega_{\mathrm{pr}}(q)=-\tfrac{i}2\Gamma+\sqrt{\Omega^2+4Dq\eta_0^{-1}\varepsilon_\infty^{-1}-\tfrac14\Gamma^2}$ is the pristine edge mode dispersion, shown in Fig.~\ref{fig:Field_B=0}(a,b,c) by dash-dotted lines and obtained from Eq.~(\ref{dispersion_B=0}) in the absence of substrate resonances (when $\varepsilon(\omega)=\varepsilon_{\infty}$ or $f_{j}=0$). Leaving only the resonant dependence on $\omega$ near the substrate phonon resonance $\omega_\mathrm{P}$, we obtain the equation written in matrix form as 
\begin{equation}\label{matrix}
\begin{vmatrix}   \omega-\omega_\mathrm{pr}(q) & \omega_{\mathrm{P}}\sqrt{f/4\varepsilon_{\infty}} \\   \omega_{\mathrm{P}}\sqrt{f/4\varepsilon_{\infty}} &  \omega-\omega_{\mathrm{P}}   \end{vmatrix} =0.
\end{equation} 
This equation for coupled pristine edge mode and substrate phonon is quite accurate for weak coupling, $\sqrt{f/4\varepsilon_{\infty}}<1$. Note that this model is applicable for EMPs too (see below).

To illustrate the anti-crossing behavior, in Fig.~\ref{fig:Field_B=0}(e) we plot the dispersion of edge modes as at varying frequency $\Omega$ of the resonance in 2D material. It can be tuned by slight change of the composition of material \cite{Ferrini_1997}, the shape of metamaterial elements \cite{Wan_2016}, or by carrier doping. As seen from Fig.~\ref{fig:Field_B=0}(e), dispersion branches exhibit anti-crossing, which is well described by Eq.~(\ref{matrix}).

\textbf{Edge magnetoplasmons.} Applying a magnetic field $\mathbf{B}\propto\hat{\mathbf{z}}$ enables formation of topological edge magnetoplasmons (see Fig.~\ref{fig:Magnetoplasmons}(a)) \cite{Jin_2016} with the low-$q$ approximate dispersion \cite{Volkov1988EdgeM,Sokolik_2024}:
\begin{equation}\label{approx_MP}
    \omega=-\frac{2q\sigma_{xy}}{\varepsilon(\omega)}\log\left(\frac{2\mathfrak{e}}{q\Lambda}\right),
\end{equation}
where $\Lambda$ is the characteristic width of the strip where charge density oscillations are concentrated. The exact dispersion of EMP is found via the Wiener-Hopf method \cite{Volkov1988EdgeM,Sokolik_2024}, and to take into account substrate resonances we replace a constant $\varepsilon$ with $\varepsilon(\omega)$ from Eq.~(\ref{varepsilon}) in the full dispersion equation (see Supplement 1), although Eq.~(\ref{approx_MP}) can also be used at small $q$.

In this section we examine EMPs in three distinct systems in the quantum Hall regime: GaAs-based quantum well (QW) on crystalline lactose, and single-layer (SLG) and bilayer graphene (BLG) on SiO$_2$ substrate. We choose the magnetic fields and electron densities so that the Fermi level $E_\mathrm{F}$ is placed between adjacent Landau levels, as shown in Fig.~\ref{fig:Magnetoplasmons}(a), and the lowest-energy dipole inter-level transition across $E_\mathrm{F}$ gives rise to cyclotron resonance in $\sigma_{\alpha\beta}(\omega)$ at the frequency $\Omega$ equal to inter-Landau-level gap. Such transition is $n=0 \to n=1$ for SLG, $n=1\to n=2$ for BLG, and $n=0\to n=1$ for QW. 

As depicted in Fig.~\ref{fig:Magnetoplasmons}(b,c,d), EMP dispersions exhibit nearly linear growth at low wavevectors and anti-crossing behavior near the cyclotron (solid blue line) and substrate (dashed lines) resonances, which in the latter case may be described by the same coupled-oscillator model (thick dashed line in Fig.~\ref{fig:Magnetoplasmons}(d)) we constructed earlier (\ref{matrix}). The anti-crossing effect caused by lactose phonons is weak due to small $f_{j}<0.05$, and leaves the pristine EMP dispersions $\omega_\mathrm{pr}(q)$ (dash-dotted lines in Fig.~\ref{fig:Magnetoplasmons}(d)) mostly untouched, deforming the curve only in the vicinity of each $\omega_{\mathrm{P},j}$. In contrast, the influence of the SiO$_2$ substrate is so pronounced ($f_{j}\sim 1$) that the dispersion curve bears little resemblance to that of a pristine EMP (Fig.~\ref{fig:Magnetoplasmons}(b,c)).

Near the substrate resonance frequencies, EMP inherits characteristics of the corresponding resonances. In particular, at $\mathrm{Re}\,\omega\approx\omega_{\mathrm{P},j}$, the quality factor, defined as $Q=\mathrm{Re}\,\omega/2\,\mathrm{Im}\,\omega$, approaches %twice
the quality factors of the phonons $\omega_{\mathrm{P},j}/\Gamma_{\mathrm{P},j}$, as shown in Fig.~\ref{fig:Magnetoplasmons}(e). Conversely, far from this resonance, the EMP quality factor is close to that of pristine EMP (calculated with $f_j=0$).

In contrast to the case without magnetic field, where the size of potential distribution in space was scaled as $q^{-1}$, for EMPs it is not the case: as may be seen from insets of Fig.~\ref{fig:Magnetoplasmons}(d), as $\omega$ rises, the field distribution plotted in the $(qx,qz)$ plane spreads out. In Fig.~\ref{fig:Magnetoplasmons}(f) we plotted the confinement factor $\lambda_{0}/(l^{\mathrm{in}}_{x}+l^{\mathrm{out}}_{x})$, which characterizes the strength of lateral field localization relative to the excitation wavelength in vacuum: it vanishes as $\propto q$ at small wavelengths for hard branches ($\omega\to\omega_{\mathrm{P},j}$) and stays constant for the soft branch ($\omega\propto q\ln(2\mathfrak{e}/q\Lambda)$).

\textbf{Conclusion.} We have demonstrated how edge phonon modes and edge magnetoplasmons hybridize with substrate resonances in the non-retarded limit. For hybrid phonon-phonon modes, we considered several polar-phonon laterally bounded thin films on polar-phonon substrates. For hybrid EMPs, we considered QW, SLG and BLG in strong magnetic fields on lactose and SiO$_{2}$ substrates. We predict prominent anti-crossing behavior of EMPs and edge modes near each substrate resonance. At realistic parameters, the quality factor of hybrid modes is of the order of 20-200, and the lengths of field confinement near the edge are 20-100 times smaller than the wavelength in free space. Our approach is valid away from the light cone, $q\gg\omega/c$, where retardation effects are negligible, and for film thicknesses and edge inhomogeneities much smaller than the wavelength $2\pi/q$. Also we assume that $\sigma(\omega)$ and $\varepsilon(\omega)$ are dominated by sharp resonant terms.

The predicted hybrid modes combine the features of pristine edge mode (e.g., strong confinement and unidirectional propagation in magnetic field) and of the bulk substrate modes (which could be phonon, exciton, plasmon etc.). Sometimes they demonstrate negative group velocity. Possible avenues of further research may be taking into account retardation, finite thickness of the film, and anisotropy. Robustness of edge magnetoplasmon topological protection \cite{Jin_2016} against hybridization with substrate resonances also deserves separate study. The edge-bulk hybridization can be used in new methods of tuning and controlling edge excitations, connecting them with bulk, surface, or waveguide modes, thereby providing opportunities for the development of new THz radiation sources, as well as novel sensing and information transmitting technologies at micro- and nanoscale.

\textbf{Funding} Foundation for the Advancement of Theoretical Physics and Mathematics (24-1-5-113-1); Russian Science Foundation (23-12-0011).

\textbf{Acknowledgment} We extend our gratitude to Oleg V. Kotov for his valuable comments and suggestions. Work on analytical Wiener-Hopf calculations was supported by the Program of Basic Research of the Higher School of Economics.

\textbf{Disclosures} The authors declare no conflicts of interest. 

\textbf{Data availability} Data underlying the results presented in this paper are not publicly available at this time but may be obtained from the authors upon reasonable request.

\textbf{Supplemental document} See Supplement 1 for supporting content.

\bibliography{References}

%apsrev4-2.bst 2019-01-14 (MD) hand-edited version of apsrev4-1.bst
%Control: key (0)
%Control: author (8) initials jnrlst
%Control: editor formatted (1) identically to author
%Control: production of article title (0) allowed
%Control: page (0) single
%Control: year (1) truncated
%Control: production of eprint (0) enabled
\begin{thebibliography}{28}%
\makeatletter
\providecommand \@ifxundefined [1]{%
 \@ifx{#1\undefined}
}%
\providecommand \@ifnum [1]{%
 \ifnum #1\expandafter \@firstoftwo
 \else \expandafter \@secondoftwo
 \fi
}%
\providecommand \@ifx [1]{%
 \ifx #1\expandafter \@firstoftwo
 \else \expandafter \@secondoftwo
 \fi
}%
\providecommand \natexlab [1]{#1}%
\providecommand \enquote  [1]{``#1''}%
\providecommand \bibnamefont  [1]{#1}%
\providecommand \bibfnamefont [1]{#1}%
\providecommand \citenamefont [1]{#1}%
\providecommand \href@noop [0]{\@secondoftwo}%
\providecommand \href [0]{\begingroup \@sanitize@url \@href}%
\providecommand \@href[1]{\@@startlink{#1}\@@href}%
\providecommand \@@href[1]{\endgroup#1\@@endlink}%
\providecommand \@sanitize@url [0]{\catcode `\\12\catcode `\$12\catcode
  `\&12\catcode `\#12\catcode `\^12\catcode `\_12\catcode `\%12\relax}%
\providecommand \@@startlink[1]{}%
\providecommand \@@endlink[0]{}%
\providecommand \url  [0]{\begingroup\@sanitize@url \@url }%
\providecommand \@url [1]{\endgroup\@href {#1}{\urlprefix }}%
\providecommand \urlprefix  [0]{URL }%
\providecommand \Eprint [0]{\href }%
\providecommand \doibase [0]{https://doi.org/}%
\providecommand \selectlanguage [0]{\@gobble}%
\providecommand \bibinfo  [0]{\@secondoftwo}%
\providecommand \bibfield  [0]{\@secondoftwo}%
\providecommand \translation [1]{[#1]}%
\providecommand \BibitemOpen [0]{}%
\providecommand \bibitemStop [0]{}%
\providecommand \bibitemNoStop [0]{.\EOS\space}%
\providecommand \EOS [0]{\spacefactor3000\relax}%
\providecommand \BibitemShut  [1]{\csname bibitem#1\endcsname}%
\let\auto@bib@innerbib\@empty
%</preamble>
\bibitem [{\citenamefont {Huang}\ \emph {et~al.}(2016)\citenamefont {Huang},
  \citenamefont {Song}, \citenamefont {Zhang},\ and\ \citenamefont
  {Yan}}]{Huang_2016}%
  \BibitemOpen
  \bibfield  {author} {\bibinfo {author} {\bibfnamefont {S.}~\bibnamefont
  {Huang}}, \bibinfo {author} {\bibfnamefont {C.}~\bibnamefont {Song}},
  \bibinfo {author} {\bibfnamefont {G.}~\bibnamefont {Zhang}},\ and\ \bibinfo
  {author} {\bibfnamefont {H.}~\bibnamefont {Yan}},\ }\bibfield  {title}
  {\bibinfo {title} {Graphene plasmonics: physics and potential applications},\
  }\href {https://doi.org/10.1515/nanoph-2016-0126} {\bibfield  {journal}
  {\bibinfo  {journal} {Nanophotonics}\ }\textbf {\bibinfo {volume} {6}},\
  \bibinfo {pages} {1191} (\bibinfo {year} {2016})}\BibitemShut {NoStop}%
\bibitem [{\citenamefont {Singh}\ and\ \citenamefont
  {Sarswat}(2023{\natexlab{a}})}]{SINGH2023100033}%
  \BibitemOpen
  \bibfield  {author} {\bibinfo {author} {\bibfnamefont {R.~S.}\ \bibnamefont
  {Singh}}\ and\ \bibinfo {author} {\bibfnamefont {P.~K.}\ \bibnamefont
  {Sarswat}},\ }\bibfield  {title} {\bibinfo {title} {From fundamentals to
  applications: The development of magnetoplasmonics for next-generation
  technologies},\ }\href
  {https://doi.org/https://doi.org/10.1016/j.mtelec.2023.100033} {\bibfield
  {journal} {\bibinfo  {journal} {Materials Today Electronics}\ }\textbf
  {\bibinfo {volume} {4}},\ \bibinfo {pages} {100033} (\bibinfo {year}
  {2023}{\natexlab{a}})}\BibitemShut {NoStop}%
\bibitem [{\citenamefont {Luxmoore}\ \emph {et~al.}(2014)\citenamefont
  {Luxmoore}, \citenamefont {Gan}, \citenamefont {Liu}, \citenamefont
  {Valmorra}, \citenamefont {Li}, \citenamefont {Faist},\ and\ \citenamefont
  {Nash}}]{Luxmoore_2014}%
  \BibitemOpen
  \bibfield  {author} {\bibinfo {author} {\bibfnamefont {I.~J.}\ \bibnamefont
  {Luxmoore}}, \bibinfo {author} {\bibfnamefont {C.~H.}\ \bibnamefont {Gan}},
  \bibinfo {author} {\bibfnamefont {P.~Q.}\ \bibnamefont {Liu}}, \bibinfo
  {author} {\bibfnamefont {F.}~\bibnamefont {Valmorra}}, \bibinfo {author}
  {\bibfnamefont {P.}~\bibnamefont {Li}}, \bibinfo {author} {\bibfnamefont
  {J.}~\bibnamefont {Faist}},\ and\ \bibinfo {author} {\bibfnamefont {G.~R.}\
  \bibnamefont {Nash}},\ }\bibfield  {title} {\bibinfo {title} {Strong coupling
  in the far-infrared between graphene plasmons and the surface optical phonons
  of silicon dioxide},\ }\href {https://doi.org/10.1021/ph500233s} {\bibfield
  {journal} {\bibinfo  {journal} {ACS Photonics}\ }\textbf {\bibinfo {volume}
  {1}},\ \bibinfo {pages} {1151} (\bibinfo {year} {2014})}\BibitemShut
  {NoStop}%
\bibitem [{\citenamefont {Yan}\ \emph {et~al.}(2013)\citenamefont {Yan},
  \citenamefont {Low}, \citenamefont {Zhu}, \citenamefont {Wu}, \citenamefont
  {Freitag}, \citenamefont {Li}, \citenamefont {Guinea}, \citenamefont
  {Avouris},\ and\ \citenamefont {Xia}}]{Yan_2013}%
  \BibitemOpen
  \bibfield  {author} {\bibinfo {author} {\bibfnamefont {H.}~\bibnamefont
  {Yan}}, \bibinfo {author} {\bibfnamefont {T.}~\bibnamefont {Low}}, \bibinfo
  {author} {\bibfnamefont {W.}~\bibnamefont {Zhu}}, \bibinfo {author}
  {\bibfnamefont {Y.}~\bibnamefont {Wu}}, \bibinfo {author} {\bibfnamefont
  {M.}~\bibnamefont {Freitag}}, \bibinfo {author} {\bibfnamefont
  {X.}~\bibnamefont {Li}}, \bibinfo {author} {\bibfnamefont {F.}~\bibnamefont
  {Guinea}}, \bibinfo {author} {\bibfnamefont {P.}~\bibnamefont {Avouris}},\
  and\ \bibinfo {author} {\bibfnamefont {F.}~\bibnamefont {Xia}},\ }\bibfield
  {title} {\bibinfo {title} {Damping pathways of mid-infrared plasmons in
  graphene nanostructures},\ }\href {https://doi.org/10.1038/nphoton.2013.57}
  {\bibfield  {journal} {\bibinfo  {journal} {Nature Photonics}\ }\textbf
  {\bibinfo {volume} {7}},\ \bibinfo {pages} {394} (\bibinfo {year}
  {2013})}\BibitemShut {NoStop}%
\bibitem [{\citenamefont {Koch}\ \emph {et~al.}(2010)\citenamefont {Koch},
  \citenamefont {Seyller},\ and\ \citenamefont {Schaefer}}]{Koch_2010}%
  \BibitemOpen
  \bibfield  {author} {\bibinfo {author} {\bibfnamefont {R.~J.}\ \bibnamefont
  {Koch}}, \bibinfo {author} {\bibfnamefont {T.}~\bibnamefont {Seyller}},\ and\
  \bibinfo {author} {\bibfnamefont {J.~A.}\ \bibnamefont {Schaefer}},\
  }\bibfield  {title} {\bibinfo {title} {Strong phonon-plasmon coupled modes in
  the graphene/silicon carbide heterosystem},\ }\href
  {https://doi.org/10.1103/physrevb.82.201413} {\bibfield  {journal} {\bibinfo
  {journal} {Physical Review B}\ }\textbf {\bibinfo {volume} {82}},\ \bibinfo
  {pages} {201413} (\bibinfo {year} {2010})}\BibitemShut {NoStop}%
\bibitem [{\citenamefont {Huang}\ \emph {et~al.}(2023)\citenamefont {Huang},
  \citenamefont {Deng}, \citenamefont {Ye}, \citenamefont {Fu}, \citenamefont
  {Zhang},\ and\ \citenamefont {Li}}]{Huang_2023}%
  \BibitemOpen
  \bibfield  {author} {\bibinfo {author} {\bibfnamefont {J.}~\bibnamefont
  {Huang}}, \bibinfo {author} {\bibfnamefont {F.}~\bibnamefont {Deng}},
  \bibinfo {author} {\bibfnamefont {F.}~\bibnamefont {Ye}}, \bibinfo {author}
  {\bibfnamefont {H.}~\bibnamefont {Fu}}, \bibinfo {author} {\bibfnamefont
  {S.}~\bibnamefont {Zhang}},\ and\ \bibinfo {author} {\bibfnamefont
  {Q.}~\bibnamefont {Li}},\ }\bibfield  {title} {\bibinfo {title} {Strong
  plasmon-phonon coupling for graphene/{hBN} thermal emitter atomic system},\
  }\href {https://doi.org/10.1016/j.carbon.2023.118210} {\bibfield  {journal}
  {\bibinfo  {journal} {Carbon}\ }\textbf {\bibinfo {volume} {213}},\ \bibinfo
  {pages} {118210} (\bibinfo {year} {2023})}\BibitemShut {NoStop}%
\bibitem [{\citenamefont {Wehmeier}\ \emph {et~al.}(2024)\citenamefont
  {Wehmeier}, \citenamefont {Xu}, \citenamefont {Mayer}, \citenamefont
  {Vermilyea}, \citenamefont {Tsuneto}, \citenamefont {Dapolito}, \citenamefont
  {Pu}, \citenamefont {Du}, \citenamefont {Chen}, \citenamefont {Zheng},
  \citenamefont {Jing}, \citenamefont {Zhou}, \citenamefont {Watanabe},
  \citenamefont {Taniguchi}, \citenamefont {Gozar}, \citenamefont {Li},
  \citenamefont {Kuzmenko}, \citenamefont {Carr}, \citenamefont {Du},
  \citenamefont {Fogler}, \citenamefont {Basov},\ and\ \citenamefont
  {Liu}}]{Wehmeier_2024}%
  \BibitemOpen
  \bibfield  {author} {\bibinfo {author} {\bibfnamefont {L.}~\bibnamefont
  {Wehmeier}}, \bibinfo {author} {\bibfnamefont {S.}~\bibnamefont {Xu}},
  \bibinfo {author} {\bibfnamefont {R.~A.}\ \bibnamefont {Mayer}}, \bibinfo
  {author} {\bibfnamefont {B.}~\bibnamefont {Vermilyea}}, \bibinfo {author}
  {\bibfnamefont {M.}~\bibnamefont {Tsuneto}}, \bibinfo {author} {\bibfnamefont
  {M.}~\bibnamefont {Dapolito}}, \bibinfo {author} {\bibfnamefont
  {R.}~\bibnamefont {Pu}}, \bibinfo {author} {\bibfnamefont {Z.}~\bibnamefont
  {Du}}, \bibinfo {author} {\bibfnamefont {X.}~\bibnamefont {Chen}}, \bibinfo
  {author} {\bibfnamefont {W.}~\bibnamefont {Zheng}}, \bibinfo {author}
  {\bibfnamefont {R.}~\bibnamefont {Jing}}, \bibinfo {author} {\bibfnamefont
  {Z.}~\bibnamefont {Zhou}}, \bibinfo {author} {\bibfnamefont {K.}~\bibnamefont
  {Watanabe}}, \bibinfo {author} {\bibfnamefont {T.}~\bibnamefont {Taniguchi}},
  \bibinfo {author} {\bibfnamefont {A.}~\bibnamefont {Gozar}}, \bibinfo
  {author} {\bibfnamefont {Q.}~\bibnamefont {Li}}, \bibinfo {author}
  {\bibfnamefont {A.~B.}\ \bibnamefont {Kuzmenko}}, \bibinfo {author}
  {\bibfnamefont {G.~L.}\ \bibnamefont {Carr}}, \bibinfo {author}
  {\bibfnamefont {X.}~\bibnamefont {Du}}, \bibinfo {author} {\bibfnamefont
  {M.~M.}\ \bibnamefont {Fogler}}, \bibinfo {author} {\bibfnamefont {D.~N.}\
  \bibnamefont {Basov}},\ and\ \bibinfo {author} {\bibfnamefont
  {M.}~\bibnamefont {Liu}},\ }\bibfield  {title} {\bibinfo {title}
  {Landau-phonon polaritons in {Dirac} heterostructures},\ }\href
  {https://doi.org/10.1126/sciadv.adp3487} {\bibfield  {journal} {\bibinfo
  {journal} {Science Advances}\ }\textbf {\bibinfo {volume} {10}},\ \bibinfo
  {pages} {eadp3487} (\bibinfo {year} {2024})}\BibitemShut {NoStop}%
\bibitem [{\citenamefont {Yan}\ \emph {et~al.}(2014)\citenamefont {Yan},
  \citenamefont {Low}, \citenamefont {Guinea}, \citenamefont {Xia},\ and\
  \citenamefont {Avouris}}]{Yan_2014}%
  \BibitemOpen
  \bibfield  {author} {\bibinfo {author} {\bibfnamefont {H.}~\bibnamefont
  {Yan}}, \bibinfo {author} {\bibfnamefont {T.}~\bibnamefont {Low}}, \bibinfo
  {author} {\bibfnamefont {F.}~\bibnamefont {Guinea}}, \bibinfo {author}
  {\bibfnamefont {F.}~\bibnamefont {Xia}},\ and\ \bibinfo {author}
  {\bibfnamefont {P.}~\bibnamefont {Avouris}},\ }\bibfield  {title} {\bibinfo
  {title} {Tunable phonon-induced transparency in bilayer graphene
  nanoribbons},\ }\href {https://doi.org/10.1021/nl501628x} {\bibfield
  {journal} {\bibinfo  {journal} {Nano Letters}\ }\textbf {\bibinfo {volume}
  {14}},\ \bibinfo {pages} {4581} (\bibinfo {year} {2014})}\BibitemShut
  {NoStop}%
\bibitem [{\citenamefont {Bryxin}\ \emph {et~al.}(1972)\citenamefont {Bryxin},
  \citenamefont {Mirlin},\ and\ \citenamefont {Reshina}}]{Bryxin_1972}%
  \BibitemOpen
  \bibfield  {author} {\bibinfo {author} {\bibfnamefont {V.}~\bibnamefont
  {Bryxin}}, \bibinfo {author} {\bibfnamefont {D.}~\bibnamefont {Mirlin}},\
  and\ \bibinfo {author} {\bibfnamefont {I.}~\bibnamefont {Reshina}},\
  }\bibfield  {title} {\bibinfo {title} {Surface plasmon-phonon interaction in
  {n-InSb}},\ }\href {https://doi.org/10.1016/0038-1098(72)90489-9} {\bibfield
  {journal} {\bibinfo  {journal} {Solid State Communications}\ }\textbf
  {\bibinfo {volume} {11}},\ \bibinfo {pages} {695} (\bibinfo {year}
  {1972})}\BibitemShut {NoStop}%
\bibitem [{\citenamefont {Greffet}\ \emph {et~al.}(2002)\citenamefont
  {Greffet}, \citenamefont {Carminati}, \citenamefont {Joulain}, \citenamefont
  {Mulet}, \citenamefont {Mainguy},\ and\ \citenamefont {Chen}}]{Greffet_2002}%
  \BibitemOpen
  \bibfield  {author} {\bibinfo {author} {\bibfnamefont {J.-J.}\ \bibnamefont
  {Greffet}}, \bibinfo {author} {\bibfnamefont {R.}~\bibnamefont {Carminati}},
  \bibinfo {author} {\bibfnamefont {K.}~\bibnamefont {Joulain}}, \bibinfo
  {author} {\bibfnamefont {J.-P.}\ \bibnamefont {Mulet}}, \bibinfo {author}
  {\bibfnamefont {S.}~\bibnamefont {Mainguy}},\ and\ \bibinfo {author}
  {\bibfnamefont {Y.}~\bibnamefont {Chen}},\ }\bibfield  {title} {\bibinfo
  {title} {Coherent emission of light by thermal sources},\ }\href
  {https://doi.org/doi: 10.1038/416061a} {\bibfield  {journal} {\bibinfo
  {journal} {Nature}\ }\textbf {\bibinfo {volume} {416}},\ \bibinfo {pages}
  {61} (\bibinfo {year} {2002})}\BibitemShut {NoStop}%
\bibitem [{\citenamefont {Požela}\ \emph {et~al.}(2014)\citenamefont
  {Požela}, \citenamefont {Požela}, \citenamefont {Šilėnas}, \citenamefont
  {Širmulis}, \citenamefont {Kašalynas}, \citenamefont {Jucienė},\ and\
  \citenamefont {Venckevičius}}]{Po_ela_2014}%
  \BibitemOpen
  \bibfield  {author} {\bibinfo {author} {\bibfnamefont {J.}~\bibnamefont
  {Požela}}, \bibinfo {author} {\bibfnamefont {K.}~\bibnamefont {Požela}},
  \bibinfo {author} {\bibfnamefont {A.}~\bibnamefont {Šilėnas}}, \bibinfo
  {author} {\bibfnamefont {E.}~\bibnamefont {Širmulis}}, \bibinfo {author}
  {\bibfnamefont {I.}~\bibnamefont {Kašalynas}}, \bibinfo {author}
  {\bibfnamefont {V.}~\bibnamefont {Jucienė}},\ and\ \bibinfo {author}
  {\bibfnamefont {R.}~\bibnamefont {Venckevičius}},\ }\bibfield  {title}
  {\bibinfo {title} {Thermally stimulated 3–15 {THz} emission at
  plasmon-phonon frequencies in polar semiconductors},\ }\href
  {https://doi.org/10.1134/S106378261412015X} {\bibfield  {journal} {\bibinfo
  {journal} {Semiconductors}\ }\textbf {\bibinfo {volume} {48}},\ \bibinfo
  {pages} {1557} (\bibinfo {year} {2014})}\BibitemShut {NoStop}%
\bibitem [{\citenamefont {Melentev}\ \emph {et~al.}(2016)\citenamefont
  {Melentev}, \citenamefont {Shalygin}, \citenamefont {Vorobjev}, \citenamefont
  {Panevin}, \citenamefont {Firsov}, \citenamefont {Riuttanen}, \citenamefont
  {Suihkonen}, \citenamefont {Korotyeyev}, \citenamefont {Lyaschuk},
  \citenamefont {Kochelap},\ and\ \citenamefont {Poroshin}}]{Melentev_2016}%
  \BibitemOpen
  \bibfield  {author} {\bibinfo {author} {\bibfnamefont {G.~A.}\ \bibnamefont
  {Melentev}}, \bibinfo {author} {\bibfnamefont {V.~A.}\ \bibnamefont
  {Shalygin}}, \bibinfo {author} {\bibfnamefont {L.~E.}\ \bibnamefont
  {Vorobjev}}, \bibinfo {author} {\bibfnamefont {V.~Y.}\ \bibnamefont
  {Panevin}}, \bibinfo {author} {\bibfnamefont {D.~A.}\ \bibnamefont {Firsov}},
  \bibinfo {author} {\bibfnamefont {L.}~\bibnamefont {Riuttanen}}, \bibinfo
  {author} {\bibfnamefont {S.}~\bibnamefont {Suihkonen}}, \bibinfo {author}
  {\bibfnamefont {V.~V.}\ \bibnamefont {Korotyeyev}}, \bibinfo {author}
  {\bibfnamefont {Y.~M.}\ \bibnamefont {Lyaschuk}}, \bibinfo {author}
  {\bibfnamefont {V.~A.}\ \bibnamefont {Kochelap}},\ and\ \bibinfo {author}
  {\bibfnamefont {V.~N.}\ \bibnamefont {Poroshin}},\ }\bibfield  {title}
  {\bibinfo {title} {Interaction of surface plasmon polaritons in heavily doped
  {GaN} microstructures with terahertz radiation},\ }\href
  {https://doi.org/10.1063/1.4943063} {\bibfield  {journal} {\bibinfo
  {journal} {Journal of Applied Physics}\ }\textbf {\bibinfo {volume} {119}},\
  \bibinfo {pages} {093104} (\bibinfo {year} {2016})}\BibitemShut {NoStop}%
\bibitem [{\citenamefont {Mast}\ \emph {et~al.}(1985)\citenamefont {Mast},
  \citenamefont {Dahm},\ and\ \citenamefont {Fetter}}]{Mast_1985}%
  \BibitemOpen
  \bibfield  {author} {\bibinfo {author} {\bibfnamefont {D.~B.}\ \bibnamefont
  {Mast}}, \bibinfo {author} {\bibfnamefont {A.~J.}\ \bibnamefont {Dahm}},\
  and\ \bibinfo {author} {\bibfnamefont {A.~L.}\ \bibnamefont {Fetter}},\
  }\bibfield  {title} {\bibinfo {title} {Observation of bulk and edge
  magnetoplasmons in a two-dimensional electron fluid},\ }\href
  {https://doi.org/https://doi.org/10.1103/PhysRevLett.54.1706} {\bibfield
  {journal} {\bibinfo  {journal} {Physical Review Letters}\ }\textbf {\bibinfo
  {volume} {54}},\ \bibinfo {pages} {1706} (\bibinfo {year}
  {1985})}\BibitemShut {NoStop}%
\bibitem [{\citenamefont {Volkov}\ and\ \citenamefont
  {Mikhailov}(1988)}]{Volkov1988EdgeM}%
  \BibitemOpen
  \bibfield  {author} {\bibinfo {author} {\bibfnamefont {V.~A.}\ \bibnamefont
  {Volkov}}\ and\ \bibinfo {author} {\bibfnamefont {S.~A.}\ \bibnamefont
  {Mikhailov}},\ }\bibfield  {title} {\bibinfo {title} {Edge magnetoplasmons:
  Low-frequency weakly damped excitations in homogeneous two-dimensional
  electron systems},\ }\href
  {https://api.semanticscholar.org/CorpusID:118820601} {\bibfield  {journal}
  {\bibinfo  {journal} {Sov. Phys. JETP}\ }\textbf {\bibinfo {volume} {67}},\
  \bibinfo {pages} {1639} (\bibinfo {year} {1988})}\BibitemShut {NoStop}%
\bibitem [{\citenamefont {Kirichek}(2001)}]{MikhailovChapter1E}%
  \BibitemOpen
  \bibfield  {author} {\bibinfo {author} {\bibfnamefont {O.}~\bibnamefont
  {Kirichek}},\ }\href {https://books.google.ru/books?id=gigbAQAAIAAJ} {\emph
  {\bibinfo {title} {Edge Excitations of Low-dimensional Charged Systems}}},\
  Horizons in world physics\ (\bibinfo  {publisher} {Nova Science},\ \bibinfo
  {address} {Hauppauge, N.Y.},\ \bibinfo {year} {2001})\BibitemShut {NoStop}%
\bibitem [{\citenamefont {Sokolik}\ and\ \citenamefont
  {Lozovik}(2024)}]{Sokolik_2024}%
  \BibitemOpen
  \bibfield  {author} {\bibinfo {author} {\bibfnamefont {A.~A.}\ \bibnamefont
  {Sokolik}}\ and\ \bibinfo {author} {\bibfnamefont {Y.~E.}\ \bibnamefont
  {Lozovik}},\ }\bibfield  {title} {\bibinfo {title} {Drift velocity of edge
  magnetoplasmons due to magnetic edge channels},\ }\href
  {https://doi.org/10.1103/physrevb.109.165430} {\bibfield  {journal} {\bibinfo
   {journal} {Physical Review B}\ }\textbf {\bibinfo {volume} {109}},\ \bibinfo
  {pages} {165430} (\bibinfo {year} {2024})}\BibitemShut {NoStop}%
\bibitem [{\citenamefont {Yan}\ \emph {et~al.}(2012)\citenamefont {Yan},
  \citenamefont {Li}, \citenamefont {Li}, \citenamefont {Zhu}, \citenamefont
  {Avouris},\ and\ \citenamefont {Xia}}]{Yan_2012}%
  \BibitemOpen
  \bibfield  {author} {\bibinfo {author} {\bibfnamefont {H.}~\bibnamefont
  {Yan}}, \bibinfo {author} {\bibfnamefont {Z.}~\bibnamefont {Li}}, \bibinfo
  {author} {\bibfnamefont {X.}~\bibnamefont {Li}}, \bibinfo {author}
  {\bibfnamefont {W.}~\bibnamefont {Zhu}}, \bibinfo {author} {\bibfnamefont
  {P.}~\bibnamefont {Avouris}},\ and\ \bibinfo {author} {\bibfnamefont
  {F.}~\bibnamefont {Xia}},\ }\bibfield  {title} {\bibinfo {title} {Infrared
  spectroscopy of tunable {Dirac} terahertz magneto-plasmons in graphene},\
  }\href {https://doi.org/https://doi.org/10.1021/nl3016335} {\bibfield
  {journal} {\bibinfo  {journal} {Nano Letters}\ }\textbf {\bibinfo {volume}
  {12}},\ \bibinfo {pages} {3766} (\bibinfo {year} {2012})}\BibitemShut
  {NoStop}%
\bibitem [{\citenamefont {Jin}\ \emph {et~al.}(2016)\citenamefont {Jin},
  \citenamefont {Lu}, \citenamefont {Wang}, \citenamefont {Fang}, \citenamefont
  {Joannopoulos}, \citenamefont {Soljačić}, \citenamefont {Fu},\ and\
  \citenamefont {Fang}}]{Jin_2016}%
  \BibitemOpen
  \bibfield  {author} {\bibinfo {author} {\bibfnamefont {D.}~\bibnamefont
  {Jin}}, \bibinfo {author} {\bibfnamefont {L.}~\bibnamefont {Lu}}, \bibinfo
  {author} {\bibfnamefont {Z.}~\bibnamefont {Wang}}, \bibinfo {author}
  {\bibfnamefont {C.}~\bibnamefont {Fang}}, \bibinfo {author} {\bibfnamefont
  {J.~D.}\ \bibnamefont {Joannopoulos}}, \bibinfo {author} {\bibfnamefont
  {M.}~\bibnamefont {Soljačić}}, \bibinfo {author} {\bibfnamefont
  {L.}~\bibnamefont {Fu}},\ and\ \bibinfo {author} {\bibfnamefont {N.~X.}\
  \bibnamefont {Fang}},\ }\bibfield  {title} {\bibinfo {title} {Topological
  magnetoplasmon},\ }\href {https://doi.org/10.1038/ncomms13486} {\bibfield
  {journal} {\bibinfo  {journal} {Nature Communications}\ }\textbf {\bibinfo
  {volume} {7}},\ \bibinfo {pages} {13486} (\bibinfo {year}
  {2016})}\BibitemShut {NoStop}%
\bibitem [{\citenamefont {Mahoney}\ \emph {et~al.}(2017)\citenamefont
  {Mahoney}, \citenamefont {Colless}, \citenamefont {Peeters}, \citenamefont
  {Pauka}, \citenamefont {Fox}, \citenamefont {Kou}, \citenamefont {Pan},
  \citenamefont {Wang}, \citenamefont {Goldhaber-Gordon},\ and\ \citenamefont
  {Reilly}}]{Mahoney_2017}%
  \BibitemOpen
  \bibfield  {author} {\bibinfo {author} {\bibfnamefont {A.~C.}\ \bibnamefont
  {Mahoney}}, \bibinfo {author} {\bibfnamefont {J.~I.}\ \bibnamefont
  {Colless}}, \bibinfo {author} {\bibfnamefont {L.}~\bibnamefont {Peeters}},
  \bibinfo {author} {\bibfnamefont {S.~J.}\ \bibnamefont {Pauka}}, \bibinfo
  {author} {\bibfnamefont {E.~J.}\ \bibnamefont {Fox}}, \bibinfo {author}
  {\bibfnamefont {X.}~\bibnamefont {Kou}}, \bibinfo {author} {\bibfnamefont
  {L.}~\bibnamefont {Pan}}, \bibinfo {author} {\bibfnamefont {K.~L.}\
  \bibnamefont {Wang}}, \bibinfo {author} {\bibfnamefont {D.}~\bibnamefont
  {Goldhaber-Gordon}},\ and\ \bibinfo {author} {\bibfnamefont {D.~J.}\
  \bibnamefont {Reilly}},\ }\bibfield  {title} {\bibinfo {title} {Zero-field
  edge plasmons in a magnetic topological insulator},\ }\href
  {https://doi.org/10.1038/s41467-017-01984-5} {\bibfield  {journal} {\bibinfo
  {journal} {Nature Communications}\ }\textbf {\bibinfo {volume} {8}},\
  \bibinfo {pages} {1836} (\bibinfo {year} {2017})}\BibitemShut {NoStop}%
\bibitem [{\citenamefont {Xiao}\ \emph {et~al.}(2024)\citenamefont {Xiao},
  \citenamefont {Xie},\ and\ \citenamefont {Wang}}]{Xiao_2024}%
  \BibitemOpen
  \bibfield  {author} {\bibinfo {author} {\bibfnamefont {D.-Y.}\ \bibnamefont
  {Xiao}}, \bibinfo {author} {\bibfnamefont {Y.-X.}\ \bibnamefont {Xie}},\ and\
  \bibinfo {author} {\bibfnamefont {Y.-S.}\ \bibnamefont {Wang}},\ }\bibfield
  {title} {\bibinfo {title} {Topological one-way edge states in locally
  resonant metamaterials},\ }\href {https://doi.org/10.1063/5.0234849}
  {\bibfield  {journal} {\bibinfo  {journal} {Journal of Applied Physics}\
  }\textbf {\bibinfo {volume} {136}},\ \bibinfo {pages} {195104} (\bibinfo
  {year} {2024})}\BibitemShut {NoStop}%
\bibitem [{\citenamefont {Song}\ and\ \citenamefont
  {Rudner}(2016)}]{Song_2016}%
  \BibitemOpen
  \bibfield  {author} {\bibinfo {author} {\bibfnamefont {J.~C.~W.}\
  \bibnamefont {Song}}\ and\ \bibinfo {author} {\bibfnamefont {M.~S.}\
  \bibnamefont {Rudner}},\ }\bibfield  {title} {\bibinfo {title} {Chiral
  plasmons without magnetic field},\ }\href
  {https://doi.org/10.1073/pnas.1519086113} {\bibfield  {journal} {\bibinfo
  {journal} {Proceedings of the National Academy of Sciences}\ }\textbf
  {\bibinfo {volume} {113}},\ \bibinfo {pages} {4658} (\bibinfo {year}
  {2016})}\BibitemShut {NoStop}%
\bibitem [{\citenamefont {Singh}\ and\ \citenamefont
  {Sarswat}(2023{\natexlab{b}})}]{Singh_2023}%
  \BibitemOpen
  \bibfield  {author} {\bibinfo {author} {\bibfnamefont {R.~S.}\ \bibnamefont
  {Singh}}\ and\ \bibinfo {author} {\bibfnamefont {P.~K.}\ \bibnamefont
  {Sarswat}},\ }\bibfield  {title} {\bibinfo {title} {From fundamentals to
  applications: The development of magnetoplasmonics for next-generation
  technologies},\ }\href {https://doi.org/10.1016/j.mtelec.2023.100033}
  {\bibfield  {journal} {\bibinfo  {journal} {Materials Today Electronics}\
  }\textbf {\bibinfo {volume} {4}},\ \bibinfo {pages} {100033} (\bibinfo {year}
  {2023}{\natexlab{b}})}\BibitemShut {NoStop}%
\bibitem [{\citenamefont {Fischetti}\ \emph {et~al.}(2001)\citenamefont
  {Fischetti}, \citenamefont {Neumayer},\ and\ \citenamefont
  {Cartier}}]{Fischetti_2001}%
  \BibitemOpen
  \bibfield  {author} {\bibinfo {author} {\bibfnamefont {M.~V.}\ \bibnamefont
  {Fischetti}}, \bibinfo {author} {\bibfnamefont {D.~A.}\ \bibnamefont
  {Neumayer}},\ and\ \bibinfo {author} {\bibfnamefont {E.~A.}\ \bibnamefont
  {Cartier}},\ }\bibfield  {title} {\bibinfo {title} {{Effective electron
  mobility in Si inversion layers in metal–oxide–semiconductor systems with
  a high-$\kappa$ insulator: The role of remote phonon scattering}},\ }\href
  {https://doi.org/10.1063/1.1405826} {\bibfield  {journal} {\bibinfo
  {journal} {Journal of Applied Physics}\ }\textbf {\bibinfo {volume} {90}},\
  \bibinfo {pages} {4587} (\bibinfo {year} {2001})}\BibitemShut {NoStop}%
\bibitem [{\citenamefont {Chaves}\ and\ \citenamefont
  {Porto}(1973)}]{Chaves_1973}%
  \BibitemOpen
  \bibfield  {author} {\bibinfo {author} {\bibfnamefont {A.}~\bibnamefont
  {Chaves}}\ and\ \bibinfo {author} {\bibfnamefont {S.}~\bibnamefont {Porto}},\
  }\bibfield  {title} {\bibinfo {title} {{Generalized Lyddane-Sachs-Teller
  relation}},\ }\href {https://doi.org/10.1016/0038-1098(73)90386-4} {\bibfield
   {journal} {\bibinfo  {journal} {Solid State Communications}\ }\textbf
  {\bibinfo {volume} {13}},\ \bibinfo {pages} {865} (\bibinfo {year}
  {1973})}\BibitemShut {NoStop}%
\bibitem [{\citenamefont {Takahashi}(1975)}]{Takahashi_1975}%
  \BibitemOpen
  \bibfield  {author} {\bibinfo {author} {\bibfnamefont {H.}~\bibnamefont
  {Takahashi}},\ }\bibfield  {title} {\bibinfo {title} {{Extension of the
  Lyddane-Sachs-Teller and Cochran-Cowley-Kurosawa relationships to a coupled
  system of many modes with damping}},\ }\href
  {https://doi.org/https://doi.org/10.1103/PhysRevB.11.1636} {\bibfield
  {journal} {\bibinfo  {journal} {Physical Review B}\ }\textbf {\bibinfo
  {volume} {11}},\ \bibinfo {pages} {1636} (\bibinfo {year}
  {1975})}\BibitemShut {NoStop}%
\bibitem [{\citenamefont {Sokolik}\ \emph {et~al.}(2021)\citenamefont
  {Sokolik}, \citenamefont {Kotov},\ and\ \citenamefont
  {Lozovik}}]{Sokolik_2021}%
  \BibitemOpen
  \bibfield  {author} {\bibinfo {author} {\bibfnamefont {A.~A.}\ \bibnamefont
  {Sokolik}}, \bibinfo {author} {\bibfnamefont {O.~V.}\ \bibnamefont {Kotov}},\
  and\ \bibinfo {author} {\bibfnamefont {Y.~E.}\ \bibnamefont {Lozovik}},\
  }\bibfield  {title} {\bibinfo {title} {Plasmonic modes at inclined edges of
  anisotropic two-dimensional materials},\ }\href
  {https://doi.org/10.1103/physrevb.103.155402} {\bibfield  {journal} {\bibinfo
   {journal} {Physical Review B}\ }\textbf {\bibinfo {volume} {103}},\ \bibinfo
  {pages} {155402} (\bibinfo {year} {2021})}\BibitemShut {NoStop}%
\bibitem [{\citenamefont {Wan}\ \emph {et~al.}(2016)\citenamefont {Wan},
  \citenamefont {Yang},\ and\ \citenamefont {Gao}}]{Wan_2016}%
  \BibitemOpen
  \bibfield  {author} {\bibinfo {author} {\bibfnamefont {W.}~\bibnamefont
  {Wan}}, \bibinfo {author} {\bibfnamefont {X.}~\bibnamefont {Yang}},\ and\
  \bibinfo {author} {\bibfnamefont {J.}~\bibnamefont {Gao}},\ }\bibfield
  {title} {\bibinfo {title} {Strong coupling between mid-infrared localized
  plasmons and phonons},\ }\href {https://doi.org/10.1364/oe.24.012367}
  {\bibfield  {journal} {\bibinfo  {journal} {Optics Express}\ }\textbf
  {\bibinfo {volume} {24}},\ \bibinfo {pages} {12367} (\bibinfo {year}
  {2016})}\BibitemShut {NoStop}%
\bibitem [{\citenamefont {Ferrini}\ \emph {et~al.}(1997)\citenamefont
  {Ferrini}, \citenamefont {Galli}, \citenamefont {Guizzetti}, \citenamefont
  {Patrini}, \citenamefont {Bosacchi}, \citenamefont {Franchi},\ and\
  \citenamefont {Magnanini}}]{Ferrini_1997}%
  \BibitemOpen
  \bibfield  {author} {\bibinfo {author} {\bibfnamefont {R.}~\bibnamefont
  {Ferrini}}, \bibinfo {author} {\bibfnamefont {M.}~\bibnamefont {Galli}},
  \bibinfo {author} {\bibfnamefont {G.}~\bibnamefont {Guizzetti}}, \bibinfo
  {author} {\bibfnamefont {M.}~\bibnamefont {Patrini}}, \bibinfo {author}
  {\bibfnamefont {A.}~\bibnamefont {Bosacchi}}, \bibinfo {author}
  {\bibfnamefont {S.}~\bibnamefont {Franchi}},\ and\ \bibinfo {author}
  {\bibfnamefont {R.}~\bibnamefont {Magnanini}},\ }\bibfield  {title} {\bibinfo
  {title} {Phonon response of {Al$_{x}$Ga$_{1-x}$Sb/GaSb} epitaxial layers by
  {Fourier-transform} infrared-reflectance and {Raman} spectroscopies},\ }\href
  {https://doi.org/10.1103/physrevb.56.7549} {\bibfield  {journal} {\bibinfo
  {journal} {Physical Review B}\ }\textbf {\bibinfo {volume} {56}},\ \bibinfo
  {pages} {7549} (\bibinfo {year} {1997})}\BibitemShut {NoStop}%
\end{thebibliography}%


%apsrev4-2.bst 2019-01-14 (MD) hand-edited version of apsrev4-1.bst
%Control: key (0)
%Control: author (8) initials jnrlst
%Control: editor formatted (1) identically to author
%Control: production of article title (0) allowed
%Control: page (0) single
%Control: year (1) truncated
%Control: production of eprint (0) enabled
\begin{thebibliography}{14}%
\makeatletter
\providecommand \@ifxundefined [1]{%
 \@ifx{#1\undefined}
}%
\providecommand \@ifnum [1]{%
 \ifnum #1\expandafter \@firstoftwo
 \else \expandafter \@secondoftwo
 \fi
}%
\providecommand \@ifx [1]{%
 \ifx #1\expandafter \@firstoftwo
 \else \expandafter \@secondoftwo
 \fi
}%
\providecommand \natexlab [1]{#1}%
\providecommand \enquote  [1]{``#1''}%
\providecommand \bibnamefont  [1]{#1}%
\providecommand \bibfnamefont [1]{#1}%
\providecommand \citenamefont [1]{#1}%
\providecommand \href@noop [0]{\@secondoftwo}%
\providecommand \href [0]{\begingroup \@sanitize@url \@href}%
\providecommand \@href[1]{\@@startlink{#1}\@@href}%
\providecommand \@@href[1]{\endgroup#1\@@endlink}%
\providecommand \@sanitize@url [0]{\catcode `\\12\catcode `\$12\catcode
  `\&12\catcode `\#12\catcode `\^12\catcode `\_12\catcode `\%12\relax}%
\providecommand \@@startlink[1]{}%
\providecommand \@@endlink[0]{}%
\providecommand \url  [0]{\begingroup\@sanitize@url \@url }%
\providecommand \@url [1]{\endgroup\@href {#1}{\urlprefix }}%
\providecommand \urlprefix  [0]{URL }%
\providecommand \Eprint [0]{\href }%
\providecommand \doibase [0]{https://doi.org/}%
\providecommand \selectlanguage [0]{\@gobble}%
\providecommand \bibinfo  [0]{\@secondoftwo}%
\providecommand \bibfield  [0]{\@secondoftwo}%
\providecommand \translation [1]{[#1]}%
\providecommand \BibitemOpen [0]{}%
\providecommand \bibitemStop [0]{}%
\providecommand \bibitemNoStop [0]{.\EOS\space}%
\providecommand \EOS [0]{\spacefactor3000\relax}%
\providecommand \BibitemShut  [1]{\csname bibitem#1\endcsname}%
\let\auto@bib@innerbib\@empty
%</preamble>
\bibitem [{\citenamefont {Fischetti}\ \emph {et~al.}(2001)\citenamefont
  {Fischetti}, \citenamefont {Neumayer},\ and\ \citenamefont
  {Cartier}}]{Fischetti_2001}%
  \BibitemOpen
  \bibfield  {author} {\bibinfo {author} {\bibfnamefont {M.~V.}\ \bibnamefont
  {Fischetti}}, \bibinfo {author} {\bibfnamefont {D.~A.}\ \bibnamefont
  {Neumayer}},\ and\ \bibinfo {author} {\bibfnamefont {E.~A.}\ \bibnamefont
  {Cartier}},\ }\bibfield  {title} {\bibinfo {title} {{Effective electron
  mobility in Si inversion layers in metal–oxide–semiconductor systems with
  a high-$\kappa$ insulator: The role of remote phonon scattering}},\ }\href
  {https://doi.org/10.1063/1.1405826} {\bibfield  {journal} {\bibinfo
  {journal} {Journal of Applied Physics}\ }\textbf {\bibinfo {volume} {90}},\
  \bibinfo {pages} {4587} (\bibinfo {year} {2001})}\BibitemShut {NoStop}%
\bibitem [{\citenamefont {Chaves}\ and\ \citenamefont
  {Porto}(1973)}]{Chaves_1973}%
  \BibitemOpen
  \bibfield  {author} {\bibinfo {author} {\bibfnamefont {A.}~\bibnamefont
  {Chaves}}\ and\ \bibinfo {author} {\bibfnamefont {S.}~\bibnamefont {Porto}},\
  }\bibfield  {title} {\bibinfo {title} {{Generalized Lyddane-Sachs-Teller
  relation}},\ }\href {https://doi.org/10.1016/0038-1098(73)90386-4} {\bibfield
   {journal} {\bibinfo  {journal} {Solid State Communications}\ }\textbf
  {\bibinfo {volume} {13}},\ \bibinfo {pages} {865} (\bibinfo {year}
  {1973})}\BibitemShut {NoStop}%
\bibitem [{\citenamefont {Takahashi}(1975)}]{Takahashi_1975}%
  \BibitemOpen
  \bibfield  {author} {\bibinfo {author} {\bibfnamefont {H.}~\bibnamefont
  {Takahashi}},\ }\bibfield  {title} {\bibinfo {title} {Extension of the
  {Lyddane-Sachs-Teller and Cochran-Cowley-Kurosawa} relationships to a coupled
  system of many modes with damping},\ }\href
  {https://doi.org/https://doi.org/10.1103/PhysRevB.11.1636} {\bibfield
  {journal} {\bibinfo  {journal} {Physical Review B}\ }\textbf {\bibinfo
  {volume} {11}},\ \bibinfo {pages} {1636} (\bibinfo {year}
  {1975})}\BibitemShut {NoStop}%
\bibitem [{\citenamefont {Gunde}(2000)}]{Gunde_2000}%
  \BibitemOpen
  \bibfield  {author} {\bibinfo {author} {\bibfnamefont {M.~K.}\ \bibnamefont
  {Gunde}},\ }\bibfield  {title} {\bibinfo {title} {Vibrational modes in
  amorphous silicon dioxide},\ }\href
  {https://doi.org/10.1016/s0921-4526(00)00475-0} {\bibfield  {journal}
  {\bibinfo  {journal} {Physica B: Condensed Matter}\ }\textbf {\bibinfo
  {volume} {292}},\ \bibinfo {pages} {286} (\bibinfo {year}
  {2000})}\BibitemShut {NoStop}%
\bibitem [{\citenamefont {Park}\ \emph {et~al.}(2021)\citenamefont {Park},
  \citenamefont {Parker‐Jervis},\ and\ \citenamefont
  {Cunningham}}]{Park_2021}%
  \BibitemOpen
  \bibfield  {author} {\bibinfo {author} {\bibfnamefont {S.~J.}\ \bibnamefont
  {Park}}, \bibinfo {author} {\bibfnamefont {R.~S.}\ \bibnamefont
  {Parker‐Jervis}},\ and\ \bibinfo {author} {\bibfnamefont {J.~E.}\
  \bibnamefont {Cunningham}},\ }\bibfield  {title} {\bibinfo {title} {Enhanced
  terahertz spectral‐fingerprint detection of $\alpha$‐lactose using
  sub‐micrometer‐gap on‐chip waveguides},\ }\href
  {https://doi.org/10.1002/adts.202100428} {\bibfield  {journal} {\bibinfo
  {journal} {Advanced Theory and Simulations}\ }\textbf {\bibinfo {volume}
  {5}},\ \bibinfo {pages} {2100428} (\bibinfo {year} {2021})}\BibitemShut
  {NoStop}%
\bibitem [{\citenamefont {Lockwood}\ \emph {et~al.}(2005)\citenamefont
  {Lockwood}, \citenamefont {Yu},\ and\ \citenamefont
  {Rowell}}]{Lockwood_2005}%
  \BibitemOpen
  \bibfield  {author} {\bibinfo {author} {\bibfnamefont {D.}~\bibnamefont
  {Lockwood}}, \bibinfo {author} {\bibfnamefont {G.}~\bibnamefont {Yu}},\ and\
  \bibinfo {author} {\bibfnamefont {N.}~\bibnamefont {Rowell}},\ }\bibfield
  {title} {\bibinfo {title} {Optical phonon frequencies and damping in {AlAs,
  GaP, GaAs, InP, InAs and InSb} studied by oblique incidence infrared
  spectroscopy},\ }\href {https://doi.org/10.1016/j.ssc.2005.08.030} {\bibfield
   {journal} {\bibinfo  {journal} {Solid State Communications}\ }\textbf
  {\bibinfo {volume} {136}},\ \bibinfo {pages} {404} (\bibinfo {year}
  {2005})}\BibitemShut {NoStop}%
\bibitem [{\citenamefont {Gorbar}\ \emph {et~al.}(2012)\citenamefont {Gorbar},
  \citenamefont {Gusynin}, \citenamefont {Kuzmenko},\ and\ \citenamefont
  {Sharapov}}]{Gorbar_2012}%
  \BibitemOpen
  \bibfield  {author} {\bibinfo {author} {\bibfnamefont {E.~V.}\ \bibnamefont
  {Gorbar}}, \bibinfo {author} {\bibfnamefont {V.~P.}\ \bibnamefont {Gusynin}},
  \bibinfo {author} {\bibfnamefont {A.~B.}\ \bibnamefont {Kuzmenko}},\ and\
  \bibinfo {author} {\bibfnamefont {S.~G.}\ \bibnamefont {Sharapov}},\
  }\bibfield  {title} {\bibinfo {title} {Magneto-optical and optical probes of
  gapped ground states of bilayer graphene},\ }\href
  {https://doi.org/10.1103/physrevb.86.075414} {\bibfield  {journal} {\bibinfo
  {journal} {Physical Review B}\ }\textbf {\bibinfo {volume} {86}},\ \bibinfo
  {pages} {075414} (\bibinfo {year} {2012})}\BibitemShut {NoStop}%
\bibitem [{\citenamefont {Gusynin}\ \emph {et~al.}(2006)\citenamefont
  {Gusynin}, \citenamefont {Sharapov},\ and\ \citenamefont
  {Carbotte}}]{Gusynin_2006}%
  \BibitemOpen
  \bibfield  {author} {\bibinfo {author} {\bibfnamefont {V.~P.}\ \bibnamefont
  {Gusynin}}, \bibinfo {author} {\bibfnamefont {S.~G.}\ \bibnamefont
  {Sharapov}},\ and\ \bibinfo {author} {\bibfnamefont {J.~P.}\ \bibnamefont
  {Carbotte}},\ }\bibfield  {title} {\bibinfo {title} {Magneto-optical
  conductivity in graphene},\ }\href
  {https://doi.org/10.1088/0953-8984/19/2/026222} {\bibfield  {journal}
  {\bibinfo  {journal} {Journal of Physics: Condensed Matter}\ }\textbf
  {\bibinfo {volume} {19}},\ \bibinfo {pages} {026222} (\bibinfo {year}
  {2006})}\BibitemShut {NoStop}%
\bibitem [{\citenamefont {Ando}\ \emph {et~al.}(1982)\citenamefont {Ando},
  \citenamefont {Fowler},\ and\ \citenamefont {Stern}}]{Ando_1982}%
  \BibitemOpen
  \bibfield  {author} {\bibinfo {author} {\bibfnamefont {T.}~\bibnamefont
  {Ando}}, \bibinfo {author} {\bibfnamefont {A.~B.}\ \bibnamefont {Fowler}},\
  and\ \bibinfo {author} {\bibfnamefont {F.}~\bibnamefont {Stern}},\ }\bibfield
   {title} {\bibinfo {title} {Electronic properties of two-dimensional
  systems},\ }\href {https://doi.org/10.1103/revmodphys.54.437} {\bibfield
  {journal} {\bibinfo  {journal} {Reviews of Modern Physics}\ }\textbf
  {\bibinfo {volume} {54}},\ \bibinfo {pages} {437} (\bibinfo {year}
  {1982})}\BibitemShut {NoStop}%
\bibitem [{\citenamefont {Fetter}(1985)}]{Fetter_1985}%
  \BibitemOpen
  \bibfield  {author} {\bibinfo {author} {\bibfnamefont {A.~L.}\ \bibnamefont
  {Fetter}},\ }\bibfield  {title} {\bibinfo {title} {Edge magnetoplasmons in a
  bounded two-dimensional electron fluid},\ }\href
  {https://doi.org/10.1103/physrevb.32.7676} {\bibfield  {journal} {\bibinfo
  {journal} {Physical Review B}\ }\textbf {\bibinfo {volume} {32}},\ \bibinfo
  {pages} {7676} (\bibinfo {year} {1985})}\BibitemShut {NoStop}%
\bibitem [{\citenamefont {Abergel}\ and\ \citenamefont
  {Fal’ko}(2007)}]{Abergel_2007}%
  \BibitemOpen
  \bibfield  {author} {\bibinfo {author} {\bibfnamefont {D.~S.~L.}\
  \bibnamefont {Abergel}}\ and\ \bibinfo {author} {\bibfnamefont {V.~I.}\
  \bibnamefont {Fal’ko}},\ }\bibfield  {title} {\bibinfo {title} {Optical and
  magneto-optical far-infrared properties of bilayer graphene},\ }\href
  {https://doi.org/10.1103/physrevb.75.155430} {\bibfield  {journal} {\bibinfo
  {journal} {Physical Review B}\ }\textbf {\bibinfo {volume} {75}},\ \bibinfo
  {pages} {155430} (\bibinfo {year} {2007})}\BibitemShut {NoStop}%
\bibitem [{\citenamefont {Kirichek}(2001)}]{MikhailovChapter1E}%
  \BibitemOpen
  \bibfield  {author} {\bibinfo {author} {\bibfnamefont {O.}~\bibnamefont
  {Kirichek}},\ }\href {https://books.google.ru/books?id=gigbAQAAIAAJ} {\emph
  {\bibinfo {title} {Edge Excitations of Low-dimensional Charged Systems}}},\
  Horizons in world physics\ (\bibinfo  {publisher} {Nova Science},\ \bibinfo
  {address} {Hauppauge, N.Y.},\ \bibinfo {year} {2001})\BibitemShut {NoStop}%
\bibitem [{\citenamefont {Horng}\ \emph {et~al.}(2011)\citenamefont {Horng},
  \citenamefont {Chen}, \citenamefont {Geng}, \citenamefont {Girit},
  \citenamefont {Zhang}, \citenamefont {Hao}, \citenamefont {Bechtel},
  \citenamefont {Martin}, \citenamefont {Zettl}, \citenamefont {Crommie},
  \citenamefont {Shen},\ and\ \citenamefont {Wang}}]{Horng_2011}%
  \BibitemOpen
  \bibfield  {author} {\bibinfo {author} {\bibfnamefont {J.}~\bibnamefont
  {Horng}}, \bibinfo {author} {\bibfnamefont {C.-F.}\ \bibnamefont {Chen}},
  \bibinfo {author} {\bibfnamefont {B.}~\bibnamefont {Geng}}, \bibinfo {author}
  {\bibfnamefont {C.}~\bibnamefont {Girit}}, \bibinfo {author} {\bibfnamefont
  {Y.}~\bibnamefont {Zhang}}, \bibinfo {author} {\bibfnamefont
  {Z.}~\bibnamefont {Hao}}, \bibinfo {author} {\bibfnamefont {H.~A.}\
  \bibnamefont {Bechtel}}, \bibinfo {author} {\bibfnamefont {M.}~\bibnamefont
  {Martin}}, \bibinfo {author} {\bibfnamefont {A.}~\bibnamefont {Zettl}},
  \bibinfo {author} {\bibfnamefont {M.~F.}\ \bibnamefont {Crommie}}, \bibinfo
  {author} {\bibfnamefont {Y.~R.}\ \bibnamefont {Shen}},\ and\ \bibinfo
  {author} {\bibfnamefont {F.}~\bibnamefont {Wang}},\ }\bibfield  {title}
  {\bibinfo {title} {Drude conductivity of {Dirac} fermions in graphene},\
  }\href {https://doi.org/10.1103/physrevb.83.165113} {\bibfield  {journal}
  {\bibinfo  {journal} {Physical Review B}\ }\textbf {\bibinfo {volume} {83}},\
  \bibinfo {pages} {165113} (\bibinfo {year} {2011})}\BibitemShut {NoStop}%
\bibitem [{\citenamefont {Sokolik}\ and\ \citenamefont
  {Lozovik}(2024)}]{Sokolik_2024}%
  \BibitemOpen
  \bibfield  {author} {\bibinfo {author} {\bibfnamefont {A.~A.}\ \bibnamefont
  {Sokolik}}\ and\ \bibinfo {author} {\bibfnamefont {Y.~E.}\ \bibnamefont
  {Lozovik}},\ }\bibfield  {title} {\bibinfo {title} {Drift velocity of edge
  magnetoplasmons due to magnetic edge channels},\ }\href
  {https://doi.org/10.1103/physrevb.109.165430} {\bibfield  {journal} {\bibinfo
   {journal} {Physical Review B}\ }\textbf {\bibinfo {volume} {109}},\ \bibinfo
  {pages} {165430} (\bibinfo {year} {2024})}\BibitemShut {NoStop}%
\end{thebibliography}%

\end{document}

\renewcommand{\theequation}{S.\arabic{equation}}
\renewcommand{\thefigure}{S.\arabic{figure}}

\title{Supplementary Material for ``Hybridization of edge modes with substrate phonons''}

\author{Azat F. Aminov}
\affiliation{National Research University Higher School of Economics, 100100 Moscow, Russia}
\affiliation{Institute of Microelectronics Technology and High Purity Materials, Russian Academy of Sciences, Chernogolovka 142432, Russia}
\author{Alexey A. Sokolik}
\email{asokolik@hse.ru}
\affiliation{Institute for Spectroscopy, Russian Academy of Sciences, 108840 Troitsk, Moscow, Russia}
\affiliation{National Research University Higher School of Economics, 100100 Moscow, Russia}

\maketitle

\section{Maxwell equations for 2D material on a substrate}

In the absence of retardation effects, the Poisson equation for a 2D material located at the plane $z=0$ reads
\begin{equation}
\nabla[\varepsilon(\mathbf{r})\nabla\varphi(\mathbf{r})]=-4\pi\rho_{2\mathrm{D}}(x,y)\delta(z),
\end{equation}
where $\varphi(\mathbf{r})$ is the scalar potential, and $\rho_{2\mathrm{D}}(x,y)$ is the 2D charge density at the material. Assuming the plane-wave solutions in the $x$, $y$ directions for the potential $\varphi(\mathbf{r}) = \varphi(z,k,q)e^{i(kx+qy)}$ and the charge density $\rho_{2\mathrm{D}}(x,y)=\rho_{2\mathrm{D}}(k,q)e^{i(kx+qy)}$, we obtain:
\begin{equation}
\nabla[\varepsilon(\mathbf{r})(i\mathbf{e}_xk+i\mathbf{e}_yq)\varphi(z,k,q)e^{i(kx+qy)} ]=-4\pi\rho_{2\mathrm{D}}(k,q)e^{i(kx+qy)}\delta(z).
\end{equation}
If the dielectric function changes from $\varepsilon(\mathbf{r})=\varepsilon_{\mathrm{s}}$ below the plane ($z<0$) to $\varepsilon(\mathbf{r})=\varepsilon_{\mathrm{c}}$ above it ($z>0$), we readily obtain a solution for this equation
\begin{equation}\label{zero_solution}
    \varphi(z,k,q)=\varphi_{z=0}(k,q)e^{-\sqrt{k^2+q^2}|z|}.
\end{equation}
Thus we get the first important conclusion, that in the non-retarded regime the penetration depth of scalar potential and electric fields are identical for both the substrate and coating, irrespective of their dielectric functions. This implies that the field distribution is symmetric about the $xy$-plane. To find $\varphi_{z=0}(k,q)$, we apply the boundary conditions at $z=0$ to match the solutions across the interface:
\begin{equation}
\varepsilon_{\mathrm{c}}\partial_{z}\varphi(z,k,q)|_{z=+0} - \varepsilon_{\mathrm{s}}\partial_{z}\varphi(z,k,q)|_{z=-0} = -4\pi\rho_{2\mathrm{D}}(k,q).
\end{equation}
Using here the solution (\ref{zero_solution}), we get
\begin{equation}\label{Fourier_phi}
    \varphi_{z=0}(k,q)=\frac{4\pi\rho_{2\mathrm{D}}(k,q)}{(\varepsilon_{\mathrm{c}}+\varepsilon_{\mathrm{s}})\sqrt{k^{2}+q^{2}}}.
\end{equation}
Thus, it may be concluded that the dielectric functions of coating and substrate enter the equations in the form of mean dielectric function $\varepsilon=\frac12(\varepsilon_{\mathrm{c}}+\varepsilon_{\mathrm{s}})$, which depends on $\omega$ in our setting. 

To get the dispersion of surface modes propagating along a uniform 2D material, in addition to Eq.~(\ref{Fourier_phi}), the continuity equation has to be used, which, assuming that fields oscillate with the frequency $\omega$, has the following form
\begin{equation}
    i\omega\rho_{2\mathrm{D}}(k,q)=[k^{2}\sigma_{xx} + q^{2}\sigma_{yy}+kq(\sigma_{xy}+\sigma_{yx})]\varphi_{z=0}(k,q).
\end{equation}
In the following, we neglect the off-diagonal elements of the conductivity tensor: they vanish for an isotropic surface at $B=0$, $\sigma_{xy}=\sigma_{yx}=0$, and sum to zero in magnetic field $B>0$ perpendicular to the $xy$ plane, $\sigma_{xy}=-\sigma_{yx}$. Substituting $\varphi_{z=0}$ from Eq.~(\ref{Fourier_phi}) into the continuity equation, we get the surface mode dispersion in the form of Eq.~(4) of the main text:
\begin{equation}
     \frac{4\pi\sqrt{k_{x}^{2}+q^{2}}\sigma_{xx}(\omega)}{i\varepsilon(\omega)\omega}=2.
\end{equation}
Again, this dispersion is not affected explicitly by $\varepsilon_{\mathrm{c}}, \varepsilon_{\mathrm{s}}$, and depends only on the mean dielectric function $\varepsilon(\omega)$.

\section{Parameters for numerical calculations}
\subsection{Parameters of polar-phonon materials}

Polar materials exhibit multiple optically active phonon resonances, and for each of them two phonon modes are crucial for characterization of a substrate dielectric function $\varepsilon_\mathrm{s}(\omega)$: transverse optical (TO) modes at frequencies $\omega_{\mathrm{TO},j}$ and longitudinal optical (LO) modes at $\omega_{\mathrm{LO},j}$. The $j$-th TO mode frequency $\omega_{\mathrm{TO},j}$ determines the resonant peak of $\varepsilon_\mathrm{s}(\omega)$, and we denote this frequency as $\omega_{\mathrm{P},j}$ in the main text: $\omega_{\mathrm{P},j}=\omega_{\mathrm{TO},j}$. The LO mode frequency $\omega_{\mathrm{LO},j}$ is determined by the equation $\varepsilon_\mathrm{s}(\omega)=0$.

While the frequencies of TO phonons, which govern the conductivity resonances discussed in the main text, are readily available in the literature, finding information about their oscillator strengths $f_j$ is more challenging. However, they can be determined from other experimentally measured quantities by employing the Cochran-Cowley-Kurosawa relation \cite{Fischetti_2001, Chaves_1973, Takahashi_1975}: 
\begin{equation}
     \varepsilon_{\mathrm{s}}(\omega) = \varepsilon_{\mathrm{s},\infty} \prod_{j}\frac{\omega_{\mathrm{LO},j}^{2}-\omega^{2}-i\omega\Gamma_{\mathrm{LO},j}}{\omega_{\mathrm{TO},j}^{2}-\omega^{2}-i\omega\Gamma_{\mathrm{TO},j}},
\end{equation}
where $\Gamma_{\mathrm{TO},j}$  ($\Gamma_{\mathrm{LO},j}$) is the transverse (longitudinal) optical phonon damping rate. This relation can be converted to the form of Eq.~(1) from the main text when different phonon modes are far away from each other, and $\omega_{\mathrm{LO},j}-\omega_{\mathrm{TO},j}\gg\Gamma_{\mathrm{TO},j}$. In this case, the oscillator strength entering the Eq.~(1) is given by
\begin{equation}
    f_{j}=\varepsilon_{\mathrm{s},\infty}(\omega_{\mathrm{LO},j}^{2}-\omega_{\mathrm{TO},j}^{2})/\omega_{\mathrm{TO},j}^{2},\label{f_j}
\end{equation}
and the damping of the oscillator can be attributed to the damping of TO mode: $\Gamma_{\mathrm{P},j}=\Gamma_{\mathrm{TO},j}$. The phonon resonance oscillator strengths for the materials in Table \ref{tab:Table} were calculated using this equation, while SiO$_{2}$ and lactose oscillator strength values (Table \ref{tab:Table2}) were taken from the literature \cite{Gunde_2000, Park_2021}. The same analysis can be carried out for dielectric function $\varepsilon_\mathrm{c}(\omega)$ of a coating material, when it is present.

To obtain the two-dimensional conductivity of the thin-film (or quasi-2D) material, whose resonance is of the phonon origin as well, we assume homogeneity of its material in the $z$ direction throughout the thickness $l_{z}$: $\sigma = \sigma_{\mathrm{3D}}l_{z}$. Here the three-dimensional conductivity is related to the dielectric function as $ \sigma_{\mathrm{3D}}(\omega)=\omega[\varepsilon(\omega)-\varepsilon_{\infty}]/4\pi i $; this relation accurately describes the conductivity near the resonance and enables the use of Eq.~(3) from the main text, which proves valuable for constructing approximations such as  Eq.~(6), Eq.~(\ref{simple_approxs}) and Eq.~(\ref{q}). The strength of the conductivity resonance is governed by the conductivity weight, which in this approximation is given by $D_j=f_j\omega_{\mathrm{P},j}^{2}l_{z}/4$. Considering thin-film semiconductors, we assume their width $l_{z}=100\,\mbox{nm}$.

\begin{table}[t]
    \centering
    \caption{The parameters used in numerical calculations for semiconductors with phonon resonances \cite{Lockwood_2005}.}
\begin{tabular}{cccccc}
    \hline
    Parameter & GaAs & InAs & InP & GaP & AlAs \\
    \hline
    $\varepsilon_{\infty}$ & 10.88 & 11.7 & 9.61 & 9.09 & 8.16 \\
    $f$ & 2 & 2.63 & 2.81 & 1.91 & 1.9 \\
    $\omega_{\mathrm{TO}}/2\pi$ (THz) & 8.05 & 6.52 & 9.1 & 10.98 & 10.79 \\
    $\Gamma_{\mathrm{TO}}/2\pi$ (THz) & 0.075 & 0.259 & 0.0839 & 0.077 & 0.133 \\
    $D$ ($\mu\mbox{m} /\mbox{ps}^{2}$)  & 127.4 & 110.4 & --- & 227.8 & --- \\
    \hline
\end{tabular}
    \label{tab:Table}
\end{table}

In Table \ref{tab:Table} we list the parameters employed for approximating the substrate dielectric function (1) and thin-film material conductivity (3), which are then used to calculate the dispersions in Fig.~2.

\subsection{Parameters for EMP dispersion calculation}

In Fig.~3 we consider coupling of EMPs with phonons in SiO$_{2}$ and lactose substrates. In contrast to the semiconductors we considered above, where only one phonon resonance is profound, in these two materials there are several phonon resonances in the frequency range we are interested in, three of which are most prominent. Their properties, which we use to calculate the EMP dispersion, are listed in Table~\ref{tab:Table2}.

\begin{table}[t]
\centering
\caption{Parameters of the three most prominent TO phonon modes in lactose and SiO$_{2}$ substrates.}
\centering
\begin{tabular}{cccccc}
\hline
Parameter & Lactose \cite{Park_2021} & SiO$_{2}$ \cite{Gunde_2000} \\
\hline
$\varepsilon_{\infty}$ & 3.12 & 2.4 \\
$f_1$, $\omega_{\mathrm{TO},1}/2\pi\,\mbox{(THz)}$, $\Gamma_{\mathrm{TO},1}/2\pi\,\mbox{(GHz)}$ & 0.05, 0.53, 25.3 & 0.92, 13.37, 1.47 \\
$f_2$, $\omega_{\mathrm{TO},2}/2\pi\,\mbox{(THz)}$, $\Gamma_{\mathrm{TO},2}/2\pi\,\mbox{(GHz)}$ & 0.0045, 1.2, 47.2 & 0.08, 24.28, 31.87 \\
$f_3$, $\omega_{\mathrm{TO},3}/2\pi\,\mbox{(THz)}$, $\Gamma_{\mathrm{TO},3}/2\pi\,\mbox{(GHz)}$ & 0.036, 1.37, 52.1 & 0.66, 31.87, 2.25 \\
\hline
\end{tabular}
  \label{tab:Table2}
\end{table}

\begin{table}[t]
    \caption{Parameters used for numerical calculation in Fig.~3.}
    \label{tab:Table3}
\centering
    \begin{tabular}{cccc}
        \hline
        Parameter & QW & SLG & BLG \\
        \hline
        $\Omega/2\pi$ (THz) & 2.6 & 41.77 & 17.31 \\
        $B$ (T) & 6.2 & 21 & 16 \\
        $n_\mathrm{e}\:(10^{12}\,\mathrm{cm}^{-2})$& 0.3 & 1 & 3.1 \\
        $m^{*}_\mathrm{e}/m_\mathrm{e}$& 0.067 & --- & 0.0365 \\
        $N_{\mathrm{max}}$ & 0 & 0 & 1 \\
        $\Omega$ & $\Omega_\mathrm{c}$ & $\Omega_\mathrm{D}$ & $\sqrt{2}\Omega_\mathrm{c}$ \\
        $\Gamma/2\pi$ (THz) & 0.004&4&4\\
        \hline
    \end{tabular}
\end{table}

In our analysis, the role of 2D materials supporting EMPs is played by GaAs-based quantum well (QW), single-layer graphene (SLG), and bilayer graphene (BLG). To minimize dissipation, we assume the 2D material to be in the quantum Hall regime, where the Fermi level is positioned between the lowest-lying Landau levels, as shown in Fig.~3(a), whose numbers are $N_\mathrm{max}$ (highest occupied) and $N_\mathrm{max}+1$ (lowest unoccupied), having in mind the selection rule $\Delta |N|=\pm1$ for optical dipole transitions. The resonance frequency of the transition $N_\mathrm{max}\rightarrow N_\mathrm{max}+1$ is 
\begin{equation}\label{En_diff}
    \Omega=\frac{E_{N_{\mathrm{max}}+1}-E_{N_{\mathrm{max}}}}{\hbar},
\end{equation}
where $E_{N}$ is the energy of the $N$-th Landau level given by \cite{Gorbar_2012, Gusynin_2006, Ando_1982}:
\begin{eqnarray}\label{S10}
    E^{\mathrm{QW}}_N=\hbar\Omega_{\mathrm{c}}(N+1/2), \\
    E^{\mathrm{SLG}}_N=\hbar\Omega_{\mathrm{D}}\sqrt{N}, \\
    E^{\mathrm{BLG}}_N=\hbar\Omega_{\mathrm{c}}\sqrt{N(N-1)} 
\end{eqnarray}
(we assume $N\geq0$ for simplicity, which corresponds to the Fermi level in the conduction band and intraband transitions in the case of SLG and BLG). Here $\Omega_{\mathrm{D}}=\sqrt{2e\hbar v_{\mathrm{F}}^{2}B/c}$ is the Dirac frequency ($v_{\mathrm{F}}$ is the Fermi velocity of electrons in SLG) and $\Omega_{\mathrm{c}}=eB/m^{*}_\mathrm{e}c$ is the cyclotron frequency, with $m^{*}_\mathrm{e}$ being the effective electron mass.

The tensor of conductivity of a two-dimensional electron system at integer Landau level filling in strong magnetic field (i.e. in the quantum Hall regime), originating from the inter-Landau level transition $N_\mathrm{max}\to N_{\mathrm{max}+1}$, is given by
\begin{equation}\label{sigma_B}
    \sigma_{xx,yy}(\omega)=\frac{iD}{\pi}\frac{\omega+i\Gamma} {(\omega+i\Gamma)^{2}-\Omega^{2}},\qquad\sigma_{xy}(\omega)=-\sigma_{yx}(\omega)=\frac{D}{\pi}\frac{\Omega} {(\omega+i\Gamma)^{2}-\Omega^{2}},
\end{equation}
where $D$ is the conductivity weight of the resonance at $\omega=\Omega$, which is given by the following expressions \cite{Gorbar_2012, Gusynin_2006, Ando_1982}
\begin{eqnarray}
    D^{\mathrm{QW}} = \frac{e^{2}}{\hbar} (N_{\mathrm{max}}+1)\Omega_{\mathrm{c}},\\
    D^{\mathrm{SLG}} = \frac{e^{2}}{\hbar}\frac{\Omega_{\mathrm{D}}/2}{\sqrt{N_{\mathrm{max}}+1} - \sqrt{N_{\mathrm{max}}}},\\
    D^{\mathrm{BLG}} = \frac{e^{2}}{\hbar}\frac{2\sqrt{N_{\mathrm{max}}}\Omega_{\mathrm{c}}}{\sqrt{N_{\mathrm{max}}+1}-\sqrt{N_{\mathrm{max}}-1}}. \label{S16}
\end{eqnarray}

Since edge magnetoplasmon characteristic properties (strong confinement, small damping) are most pronounced at frequencies below the cyclotron resonance frequency $\omega<\Omega$ \cite{Fetter_1985}, we want to maximize $\Omega$ to make it higher or at least of the order of substrate phonon frequencies. As the energy differences (\ref{En_diff}) increase with the magnetic field $B$, maximizing the field is crucial for maximizing the resonance frequency. However, the magnetic field should not be too strong for at least one Landau level to be fully occupied. This constraint on the Landau level filling factor $\nu =2\pi n_\mathrm{e}\hbar c/g_\mathrm{v}g_\mathrm{s}Be$ (here $g_\mathrm{v},g_\mathrm{s}$ are the valley and spin degeneracies, $n_\mathrm{e}$ is the electron density), which has to be at least \cite{Abergel_2007} 0.5 for SLG ($N=0 \to N=1$ transition) or 1 for BLG and QW ($N=1\to N=2$ and $N=0\to N=1$ transitions respectively), restricts the maximum achievable field strength. For our numerical calculations, we assume realistic $n_{\mathrm{e}}
\sim10^{12}\,\mbox{cm}^{-2}$ and magnetic fields, which enable these transitions: 6.2~T ($N_{\mathrm{max}}=0$), 21~T ($N_{\mathrm{max}}=0$), and 16~T ($N_{\mathrm{max}}=1$) for QW, SLG, and BLG, respectively. In the case of interband transitions $-N\to N+1$ or $-N-1\to N$ in SLG and BLG magnetic fields can be lower.

Using expressions (\ref{S10})--(\ref{S16}), one can obtain the conductivity resonance frequencies and the corresponding conductivity weights for all three materials (QW, SLG, and BLG), which depend on two parameters: magnetic field $B$ and electron density $n_\mathrm{e}$. The parameters we use for calculations of EMP dispersions are listed in Table~\ref{tab:Table3}, where $m_\mathrm{e}$ is the free electron mass, and the decay rates $\Gamma$ were taken from Refs.~\cite{MikhailovChapter1E} and \cite{Horng_2011} for QW and SLG respectively. The damping rate of electrons in BLG is assumed to be the same as in SLG.

\begin{figure}[b]
    \centering
    \includegraphics[width=0.6\textwidth]{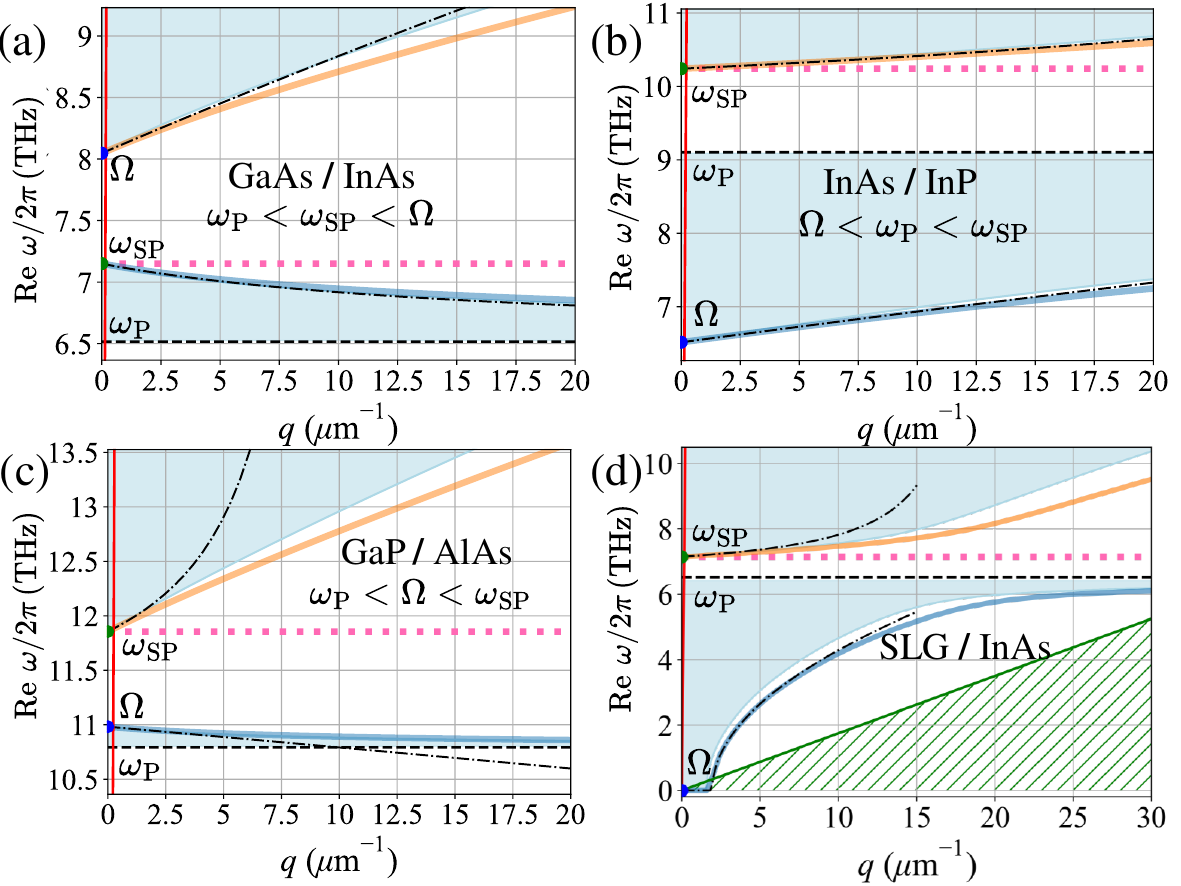}
    \caption{Dispersions of edge modes (solid lines) and their analytic approximations (\ref{simple_approxs})--(\ref{simple_approxs2}) shown by dash-dotted lines. Panels (a-c) depict the same combination of materials as in Fig.~2 from the main text with the same curve designations. Panel (d) depicts the case of edge plasmons on SLG with Drude-like metallic response (where the resonance frequency of electronic system is zero, $\Omega=0$) placed on top of the phonon-resonant substrate. Green hatching shows the continuum of intraband electron-hole excitations in graphene $\omega<v_\mathrm{F}q$.}
    \label{fig:S_edge}
\end{figure}

\section{More accurate approximation for edge mode dispersion}
Here we construct the analytical approximation for edge mode dispersions in the absence of a magnetic field for the case of a single prominent substrate resonance with the parameters $f$, $\omega_{\mathrm{P}}$, $\Gamma_{\mathrm{P}}$. This approximation is more accurate than the coupled-oscillator model, yet it lacks a clear physical interpretation.

As depicted in Fig. 2, two distinct branches can be identified. The first branch, $\omega_{1}$, is likely associated with surface phonons, as it approaches the surface phonon frequency ($\omega_1\approx\omega_{\mathrm{SP}}$) at low $q$. The second branch, $\omega_{2}$, corresponds to the resonant mode of the 2D material, as $\omega_2\approx\Omega$ in the same low-$q$ limit. Equations for $\omega_{1}(q)$ and $\omega_{2}(q)$ in this approximation are obtained by converting  biquadratic equation over $\omega$, which we reproduce from the main text (Eq.~(4)),
\begin{equation}\label{eq4}
    \frac{4\pi q\sigma_{xx}(\omega)}{i\varepsilon(\omega)\omega}=\eta_{0},
\end{equation}
to the quadratic one. Namely, to obtain an approximation for the surface phonon-like branch $\omega_{1}(q)$, in the Eq.~(\ref{eq4}) we replace $\omega$ in the denominator of the conductivity $\sigma(\omega)$ by $\omega_{\mathrm{SP}}$ and get the quadratic equation for $\omega$. Similarly, to obtain an approximation for the mode $\omega_2(q)$ preferentially attributed to the 2D material resonance, we solve Eq.~(\ref{eq4}) with $\varepsilon(\omega)$ approximated by $\varepsilon(\Omega)$. These procedures yield
\begin{eqnarray}\label{simple_approxs}
\omega_1(q) \approx -\frac{i\Gamma_{\mathrm{P}}}{2} + \sqrt{ \omega_{\mathrm{P}}^{2} + \frac{f\omega_{\mathrm{P}}^{2}}{2\left(\varepsilon_{\infty} - \frac{4Dq}{\eta_{0}(\Omega^{2}-\omega_{SP}^{2})}\right) } -\frac{\Gamma_{\mathrm{P}}^{2}}{4}}, \\
     \omega_2(q) \approx -\frac{i\Gamma}{2} +
\sqrt{\Omega^{2}+\frac{4Dq}{\eta_{0}\varepsilon(\Omega)} - \frac{\Gamma^{2}}{4}}.\label{simple_approxs2}
\end{eqnarray}
In Fig.\ref{fig:S_edge}(a,b,c), we present the edge mode dispersion for the same conditions and materials as those shown in Fig. 1 of the main text. The dash-dotted lines represent the approximations given by Eqs.~(\ref{simple_approxs}) and (\ref{simple_approxs2}), which are in close agreement with the exact solutions of Eq.~(\ref{eq4}), depicted as solid lines, when branches are far apart from each other (panel (b)), and quickly become inaccurate as $q$ rises when branches are relatively close to each other (panels (a) and (c)).

There is a peculiar case when the 2D material is a metal with the Drude-like surface conductivity ($\Omega=0$), which gives rise to edge plasmons. Due to non-zero damping of oscillations in the conductor $\Gamma$, the plasmons at low $q$ become overdamped, so their dispersion starts from the finite wavevector $q_{0}=\Gamma^{2}\eta_{0}\varepsilon(0)/16D$. As example for calculations shown in  Fig.~\ref{fig:S_edge}(d), we consider single-layer graphene whose conductivity weight (also referred to as the Drude weight) is $D=(v_{\mathrm{F}}e^{2}/\hbar) \sqrt{\pi|n_\mathrm{e}|}$ \cite{Horng_2011}, where $v_{\mathrm{F}}=1.1\times10^{8}\,\mbox{cm/s}$ is the Fermi velocity of massless electrons, $n_\mathrm{e}=10^{12}$ cm$^{-2}$ is the electron density. The damping rate is assumed to be $\Gamma/2\pi=4\,\mbox{THz}$ \cite{Horng_2011}.

\begin{figure}[b]
    \centering    \includegraphics[width=0.6\textwidth]{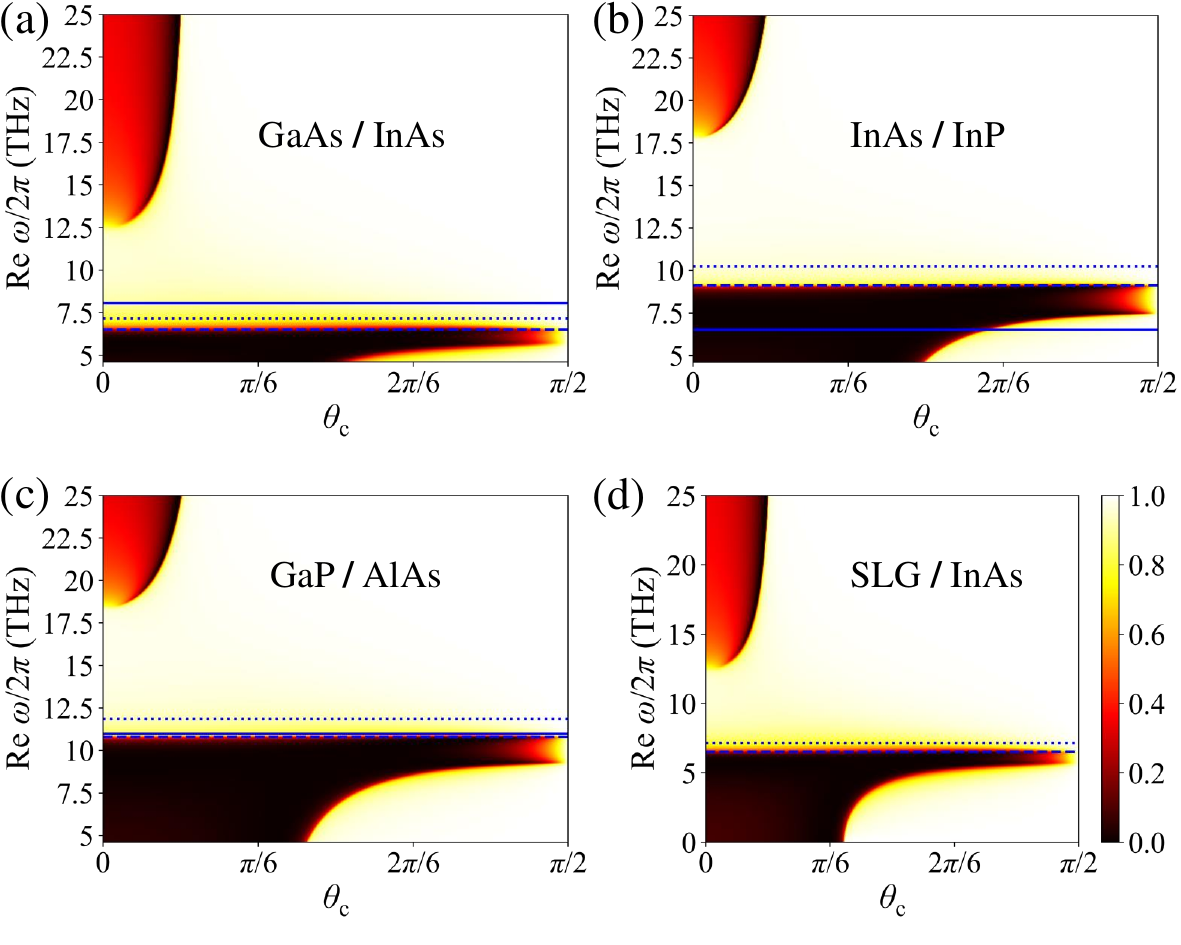}
    \caption{Intensity reflection coefficient $R_\mathrm{TM}$ for TM-polarized wave as function of incidence angle $\theta_\mathrm{c}$ and frequency $\omega$. Combinations of 2D material and substrate in panels (a)-(d) are the same as in the corresponding panels of Fig.~2. Horizontal dashed, dotted, and solid lines mark, respectively, resonant frequencies of bulk substrate phonon $\omega_\mathrm{P}$, surface substrate phonon $\omega_\mathrm{SP}$, and phonon of the 2D material $\Omega$ (for panels (a-c) only).}
    \label{fig:S_R_TM}
\end{figure}

\begin{figure}[t]
    \centering    \includegraphics[width=0.6\textwidth]{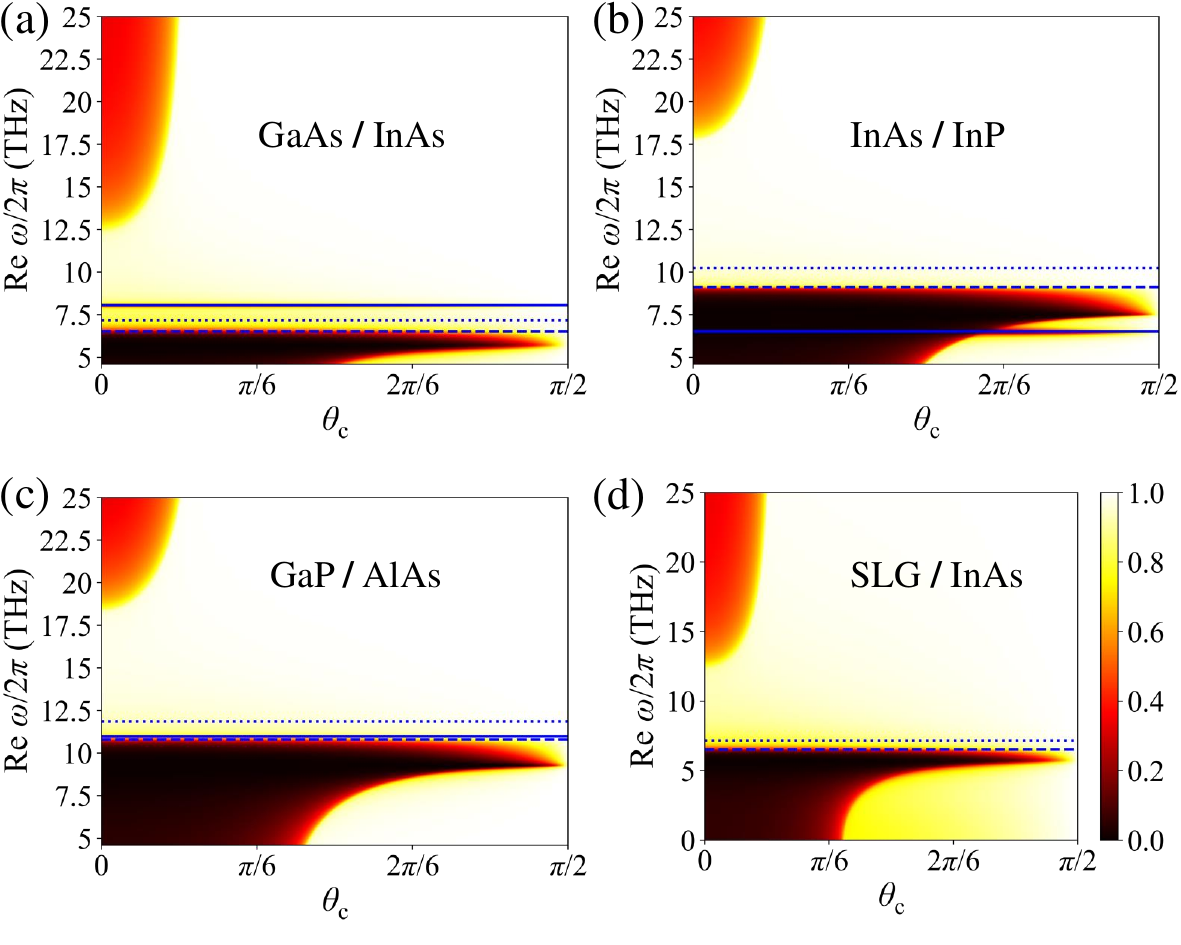}
    \caption{The same as Fig.~\ref{fig:S_R_TM} but for reflection $R_\mathrm{TE}$ of TE-polarized electromagnetic wave.}
    \label{fig:S_R_TE}
\end{figure}

\section{Reflectivity and quality factor}
In this section, we provide additional information on edge modes resulting from the hybridization of edge phonons in 2D material and bulk phonons in a substrate. Reflection coefficients for light intensity, when the light is incident from the coating medium on the substrate covered by 2D material, are given by the modified Fresnel formulas
\begin{equation}
R_\mathrm{TM}=\left|\frac{\frac{\sqrt{\varepsilon_\mathrm{c}}}{\cos\theta_\mathrm{c}}-\frac{\sqrt{\varepsilon_\mathrm{s}}}{\cos\theta_\mathrm{s}}-\frac{4\pi\sigma}c}{\frac{\sqrt{\varepsilon_\mathrm{c}}}{\cos\theta_\mathrm{c}}+\frac{\sqrt{\varepsilon_\mathrm{s}}}{\cos\theta_\mathrm{s}}+\frac{4\pi\sigma}c}\right|^2,\qquad
R_\mathrm{TE}=\left|\frac{\sqrt{\varepsilon_\mathrm{c}}\cos\theta_\mathrm{c}-\sqrt{\varepsilon_\mathrm{s}}\cos\theta_\mathrm{s}-\frac{4\pi\sigma}c}{\sqrt{\varepsilon_\mathrm{c}}\cos\theta_\mathrm{c}+\sqrt{\varepsilon_\mathrm{s}}\cos\theta_\mathrm{s}+\frac{4\pi\sigma}c}\right|^2
\end{equation}
for two light polarizations. Here $\theta_\mathrm{c}$ is the angle of incidence, $\theta_\mathrm{s}$ is the angle of refraction given by equation $\sqrt{\varepsilon_\mathrm{c}}\sin\theta_\mathrm{c}=\sqrt{\varepsilon_\mathrm{s}}\sin\theta_\mathrm{s}$. Diagrams for the reflection coefficients are shown in Figs.~\ref{fig:S_R_TM}-\ref{fig:S_R_TE} for the same combinations of 2D material and substrate as in Fig.~2 in the main text. We can notice Reststrahlenbands, i.e. bands of very strong reflection near normal incidence confined between the bulk frequency of TO substrate phonon $\omega_\mathrm{P}$ and bulk frequency of LO substrate phonon $\omega_\mathrm{P}\sqrt{1+f/\varepsilon_{s,\infty}}$ given by Eq.~(\ref{f_j}) at weak damping. Hybrid edge modes, although they exist at much higher in-plane momenta $q$ outside the light cone, mostly fit inside the Reststrahlenband frequency regions.

\begin{figure}[t]
    \centering
    \includegraphics[width=0.6\textwidth]{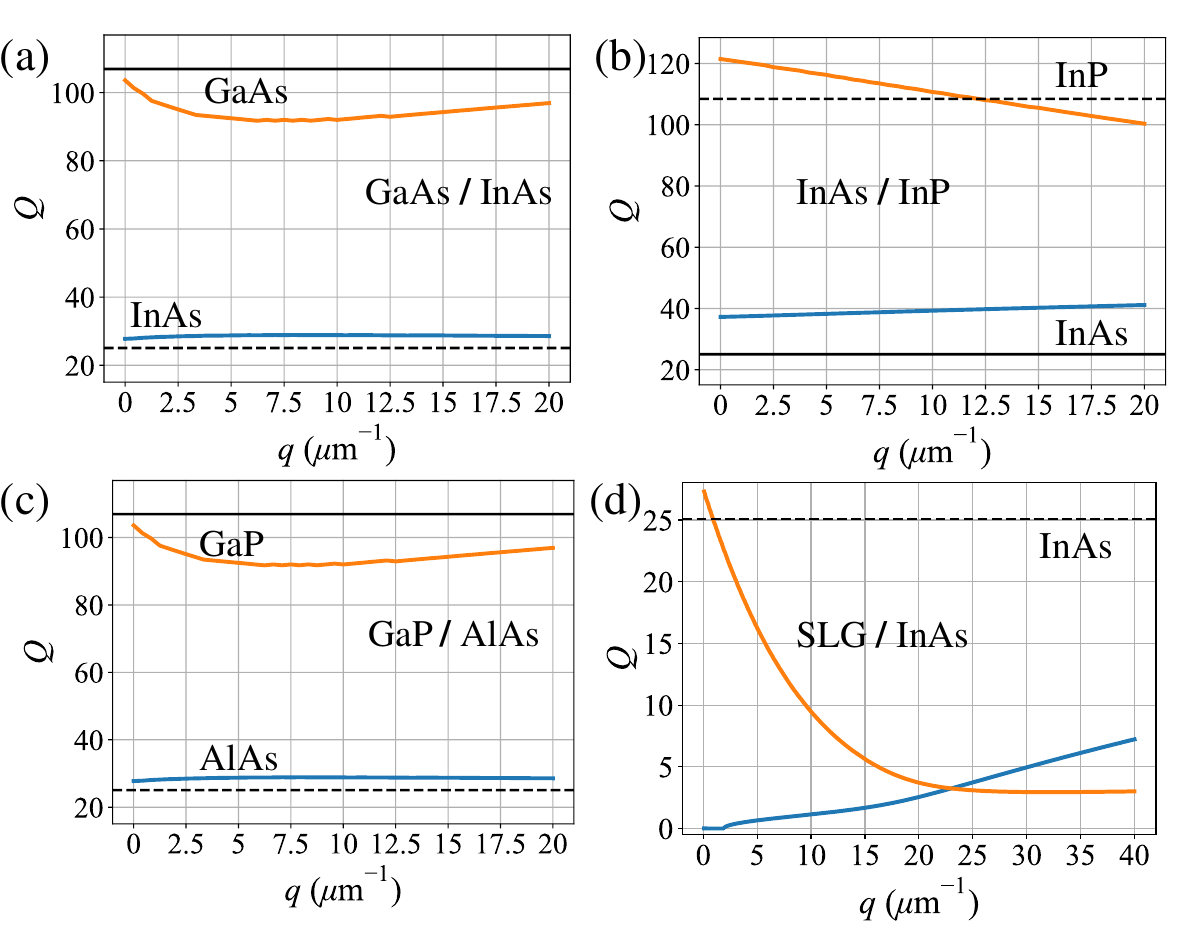}
    \caption{(a-c) Quality factors $Q=\mathrm{Re}\,\omega/2\,\mathrm{Im}\,\omega$ of lower (blue solid line) and upper (orange solid line) edge modes for the same pairs of 2D/substrate materials as in Fig.~\ref{fig:S_edge}. Black solid and dashed lines show phonon quality factors of 2D material ($\Omega/\Gamma$) and substrate ($\omega_\mathrm{P}/\Gamma_\mathrm{P}$). (d) The same for hybridization of edge plasmon in SLG with substrate phonon in GaAs. Quality factor of the lower gapless branch (blue solid line) is low due to high damping rate $\Gamma/2\pi=4\,\mbox{THz}$ in SLG, and that of the upper branch (orange solid line) is close to that of the substrate phonon (dashed line).}
    \label{fig:S_edgeQ}
\end{figure}

In Fig.~\ref{fig:S_edgeQ} we show quality factors of lower and upper edge modes in the same setups as in Fig.~\ref{fig:S_edge}, including edge plasmons in SLG on GaAs substrate. We see that edge modes inherit their quality factors from those phonon modes that are the closest in frequency (dashed and solid lines).

\section{Edge magnetoplasmon dispersion equation}

Equation governing the EMP dispersion in the non-retarded limit, which is derived using the Wiener-Hopf method, has the following analytical form \cite{Sokolik_2024}:
\begin{eqnarray}\label{EMP_eq}
\frac{1+\lambda}{1-\lambda} \exp \left\{\frac{2 i}{\pi} f(\lambda)\right\}=\frac{\chi+\eta}{\chi-\eta},\\
\lambda = -2i/\eta + \sqrt{1 - (2/\eta)^2},\label{eta_lambda}
\end{eqnarray}
where $\chi$ and $\eta$ are dimensionless functions related to the conductivities,
\begin{equation}
\eta=\frac{4 \pi q \sigma_{xx}(\omega)}{i \varepsilon(\omega) \omega}, \quad \chi=\frac{4 \pi q \sigma_{x y}(\omega)}{\varepsilon(\omega) \omega},
\end{equation}
and the function $f(z)$ is
\begin{equation}
    f(z)= -\frac{\pi^2}{6}+\ln z \ln (1+z)
+\mathrm{Li}_2(-z)+\mathrm{Li}_2(1-z),
\end{equation}
where Li$_{2}(z)$ is the dilogarithm function. Solving this equation one can obtain the complex EMP dispersion $\omega(q)$, where $\mathrm{Im}\,\omega(q)$ corresponds to EMP damping rate. Moreover, as shown in Ref.~\cite{Sokolik_2024}, the approximation for EMP dispersion, Eq.~(7) in the main text, can be derived from this equation in the low-frequency and long-wavelength limit $\omega\ll\Omega$, $|\eta|\ll1$.

It is useful to find the wavevector $q_{\omega=\Omega}$ at which the EMP dispersion intersects the cyclotron resonance frequency $\omega(q)=\Omega$ in the clean limit $\Gamma=0$ and with conductivity in the form of Eq.~(\ref{sigma_B}). In this case, Eq.~(\ref{EMP_eq}) can be rewritten as follows
\begin{equation}\label{EMP_simple}
    \frac{1+\lambda}{1-\lambda} \exp \left\{\frac{2 i}{\pi} f(\lambda)\right\}=\frac{\Omega+\omega}{\Omega-\omega}.
\end{equation}
As $\omega$ approaches $\Omega$, both right- and left-hand sides of (\ref{EMP_simple}) diverge, and thus we try to find an approximate solution as $\lambda\approx 1+ix$, where $|x|\ll1$. By neglecting the $\lambda$-dependence of the function $f(\lambda)$, which can be approximated as $f(\lambda)\approx f(1)=-\pi^{2}/4$ in (\ref{EMP_simple}), we obtain the solution $x=(\Omega-\omega)/\Omega$. Subsequently, utilizing the relationship $\eta\approx2/\mathrm{Im}\lambda$, which is applicable at small $x$ when $\lambda\approx1$, we can derive the desired wavevector 
\begin{equation}\label{q}
    q_{\omega=\Omega}=\frac{\Omega^{2}\varepsilon(\Omega)}{D}.
\end{equation}
This result does not change whether we approach $\omega=\Omega$ from above ($\omega>\Omega$) or below ($\omega<\Omega$), meaning that in the clean limit $\Gamma=0$, when conductivity is in the form of (\ref{sigma_B}), the dispersion of EMP is a continuous function near $\omega=\Omega$. This fact leads to a conclusion, that the discontinuity of EMP dispersion near the cyclotron resonance observed in Fig.~3 comes solely from the non-zero damping $\Gamma$ within the electron system.

\bibliography{RefSuppl}